\documentclass{ws-m3as}
 \usepackage[a4paper, portrait, 
 left=1.51in,
 right=1.51in,
 top=1.65in,
 bottom=1.75in]{geometry}
\usepackage{graphicx}
\usepackage{subcaption}
\graphicspath{ {images/} }
\usepackage[english]{babel}
\usepackage{amsmath}

\usepackage{amsfonts}
\usepackage{subcaption}
\usepackage{enumerate}
\usepackage{lipsum}
\usepackage[autostyle, english = american]{csquotes}
\MakeOuterQuote{"}

\newcommand{\dom}{{n}}
\usepackage[superscript]{cite}

\usepackage{cite}
 
\usepackage{mathtools}\usepackage{placeins}
\usepackage{float}
\usepackage{wrapfig}
\usepackage{amssymb, url}

\usepackage[font=small]{caption}
\usepackage{mathtools}
\usepackage{subcaption}
\usepackage[english]{babel}
\usepackage{cite}
\usepackage[utf8]{inputenc}
\makeatletter
\renewcommand{\fnum@figure}{Fig. \thefigure}
\makeatother

\begin{document}
\markboth{C. Pan, B. Li,  C. Wang,  Y. Zhang, N. Geldner,  L. Wang,  A. L. Bertozzi}{TRUNCATED L\'EVY FLIGHTS FOR RESIDENTIAL BURGLARY}

%
%

\title{\vskip - 8 mm CRIME MODELING WITH  TRUNCATED L\'EVY FLIGHTS FOR RESIDENTIAL BURGLARY MODELS}

\author{CHAOHAO PAN\footnote{Currently at Courant Institute of Mathematical Sciences, New York University, New York, NY 10012, USA, cp2780@nyu.edu.}  \footnotemark[6] }

\author{BO LI\footnote{Currently at Department of Mathematics, University of California, Berkeley, Berkeley, CA 94720, USA, boli@math.berkeley.edu.}  \footnotemark[6]  }

\author{CHUNTIAN WANG\footnote{Corresponding author.} \footnotemark[8]}


\author{YUQI ZHANG\footnote{Currently at Uber Technologies, Inc., San Francisco, CA 94103, USA, yuqiz@uber.com.}  \footnotemark[6]}


\author{NATHAN GELDNER\footnote{Currently at Centers for Disease Control and Prevention, Atlanta, GA 30341, USA, nathangeldner@gmail.com.}  \hspace{-0.55mm} \footnote{These authors were undergraduate REU students at  California Research Training Program in Computational and Applied Mathematics, University of California, Los Angeles,  Los Angeles, USA.}}


\author{LI WANG\footnote{Department of Mathematics, Computational and Data-Enabled Science and Engineering Program, The State University of New York at Buffalo, Buffalo, NY 14260, USA, lwang46@buffalo.edu.}}

\author{ANDREA L. BERTOZZI\footnote {Department of Mathematics, University of California, Los Angeles, Los Angeles, CA 90095, USA, cwang@math.ucla.edu and bertozzi@math.ucla.edu.}}

\maketitle







\begin{history}
\received{(Day Month Year)}
\revised{(Day Month Year)}
\comby{(xxxxxxxxxx)}
\end{history}

\begin{abstract}

Statistical agent-based models for crime have shown that repeat victimization can lead to   predictable crime hotspots       (see e.g. Short et  al., \textit{Math. Models Methods Appl.},  $2008$),  
then a recent study in one space dimension (Chaturapruek et  al., \textit{SIAM J. Appl. Math}, $2013$)   shows that the hotspot dynamics changes when movement patterns of the criminals  involve long-tailed L\'evy distributions for the jump length  as opposed to classical  random walks. 
  In reality  criminals move in confined areas with a maximum jump length. 
  In this paper we  develop a mean-field continuum model with    truncated L\'evy flights for   residential burglary   in one space dimension.   The continuum model    yields local Laplace diffusion,   rather than fractional diffusion.     We present an asymptotic theory to derive the continuum equations and    show excellent agreement between the continuum model  and the agent-based simulations. 
   This suggests  that local diffusion models are universal for continuum limits of this problem,  the important quantity being the diffusion coefficient.    Law enforcement agents are  also  incorporated into the model, and the relative effectiveness of their deployment strategies are compared quantitatively. 
  

\end{abstract}

\keywords{Crime models;   truncated L\'evy flights; law enforcement agents.}

\ccode{AMS Subject Classification:  35R60, 35Q84, 60G50.}

\section{Introduction}

Residential  crime   is one of the toughest    issues in   modern society.  A quantitative,  informative and applicable   model of crime     is needed to assist    law enforcement. Crimes  of opportunity often have consistent statistical properties,  and it is      possible to model them using quantitative tools.\cite{uclaModel}
In the past ten years applied mathematicians  have been working in the burgeoning area of     crime modeling and prediction (see e.g. Refs. 
  \refcite{bellomo}-\refcite{Berestycki2013},   \refcite{Gorr2014}-\refcite{Goudon},  \refcite{koloko}, \refcite{Tijana},  \refcite{lloyd},  \refcite{Ajmone}, \refcite{mccalla2013}, \refcite{mohler2011}-\refcite{uclaModel}, \refcite{wardurban}-\refcite{Zipkin}),  since     the seminal  work Ref. \refcite{uclaModel} on 
   the mathematics of agent-based models for residential burglary.  

Roughly speaking, there are two classes of burglary models.  Class I   is statistical in nature aiming to predict the patterns of observed events. 
Among them,   self-exciting point process models in Ref.  \refcite{mohler2011} have led to   the development of software products for field use \cite{mohler2015}. 
Class II   is
 agent-based  and describes the actions of individuals that lead  to aggregate pattern formation. It is this class of models that we address here.  
Agent-based models could be used for prediction if all model parameters were known. Parameters for environmental variables can be well estimated from field data, however movement patterns of individual burglars are difficult to track. Therefore, it is imperative to identify the simplest class of universal models for criminal movement. 
%

%
%

  Ref. \refcite{uclaModel} used a  biased random walk, that is,   short hops,  for criminal agents.   
It is well-known that people foraging in an environment are more likely to move according to a    \textit{L\'evy}   flight than a random walk\cite{human, travel, foraging}.   A later paper\cite{levycrime}  analyzed such processes for this model and showed that such processes  lead  to fractional diffusion rather than classical Brownian motion in the continuum. Here we refocus   the analysis to truncated \textit{L\'evy} flights since they are the most realistic.  The truncation size represents  the maximum mobility  of an agent.  
%
  We  show that an analogue mean-field continuum model exists with local diffusion replacing fractional diffusion.
 Specifically, for a range of length scales   truncated 
  \textit{L\'evy} flights behave similarly to a Brownian process with a modified diffusion coefficient. 
 
 As for the coupling of the dynamics of criminals and of the environment variables, following Ref. \refcite{uclaModel}, we   incorporate  the repeat and near-repeat victimization and the broken windows effect. These are concepts in criminology and sociology that have been empirically observed\cite{budd1999burglary, bw2, bw1}. Specifically, residential burglars prefer to return to a previously burglarized house and its neighbors\cite{repeat, residential, bw3, bw4, bw5}. These are known as repeat  and near-repeat events.  Also according to  the ``broken windows'' theory,   it is very likely that the visible signs of the past crimes in a neighborhood  may create an environment that encourages   further illegal activities \cite{wilson1982}.   
 %

  %
  %
    
    In addition following Ref. \refcite{law_enforcement},  we     introduce   the effects of   law enforcement agents into the model.   In Ref. \refcite{law_enforcement}, all    the agents  are assumed to  take      random walks, while here    law enforcement agents  follow truncated \textit{L\'evy} flights whose    maximum jump length can be different from that of the criminals. 
The relative effectiveness of several policing strategies is compared quantitatively.  

This is the first time that    truncated \textit{L\'evy} flights have been applied in crime modeling. Previously they have only  been applied in   other areas such as finance \cite{financial, option_theory, Economic} and networks.\cite{cao}

  

The article is organized as follows. In  Sec.  \ref{sec:discrete model}, we  introduce the discrete model and compare it for different values of the jump length. In  Sec.  \ref{continulimit}, we derive the mean-field continuum model   and compare it with the discrete model through computer simulations. Next 
in  Sec.  \ref{sec:3}, we incorporate law enforcement agents into the system, derive the continuum equations, and then compare the relative effectiveness of the deployment strategies   both quantitatively and qualitatively. 

\section{Discrete Model}\label{sec:discrete model}

\subsection{Overview}\label{sec:overview}
As in Ref. \refcite{levycrime},  the system is defined on a one-dimensional grid which represents  the stationary burglary sites. We assume constant grid  lattice spacing $l$ and periodic boundary conditions. Our model consists of two components --- the stationary burglary sites and a collection of burglar agents jumping from site to site. 
The   system  evolves only at   discrete time steps   $t=n\delta t, n\in \mathbb{N}$, $\delta t >0$. Attached to each grid $k\in \mathbb Z$   is a vector $(\dom_k(t), A_k (t))$,  representing the  number of criminals and   the ``attractiveness'' at site $k$ at time $t$. The  attractiveness refers to the burglar's beliefs about the vulnerability and value of the target site and it is assumed to consist of   a static background term  and a dynamic term:
\begin{equation}
{A}_k(t)=A_k^0+{B}_k(t). 
\label{eq:decomposed}
\end{equation} The dynamic term $B_k(t)$ represents the component associated with repeat victimization and broken windows effect, whose behavior  will be discussed shortly. 
Our model unfolds starting  with some initial distribution of criminal agents and attractiveness field  over the lattice grid. At   each time step,  the system gets updated as follows:

\textit{Step 1.}  Every criminal decides if he will    commit a burglary at his current site  with probability
\begin{equation}
p_k(t)=1-e^{-A_k(t)\delta t}.
\label{eq:pr}
\end{equation}  This means that the Poisson instantaneous burglary rate is roughly $A_k(t)$, and $A_k (t) \delta t$ is the expected number of burglary   events in the time interval of length $\delta t$ from a single burglar at site $k$. 

\textit{Step 2}. 
 If a criminal agent chooses to commit a burglary then he will be immediately removed from the system. Otherwise he will move to another site according to a truncated   \textit{L\'evy} distribution biased towards areas  with a high attractiveness. 
More specifically, 
the probability  of an agent jumping from site $k$ to  $i$  is
\begin{equation}
q_{k \rightarrow i} (t)= \dfrac{w_{k\rightarrow i} (t)}{\sum\limits_{\substack {j \in \mathbb Z \\ j\neq k}}{w_{k \rightarrow j}(t)}},\quad \quad k \neq i,
\label{eq:movq}
\end{equation}
where the corresponding relative weight $w_{k\rightarrow i} (t)$ is defined as
\begin{equation}
w_{k\rightarrow i} (t)=
\begin{cases} 
\dfrac{A_i (t)}{l^{\mu}|i-k|^{\mu}},
 & 1\leq |i-k| \leq L, \\
0, & \text{otherwise}.
\end{cases}
\label{eq:weight}
\end{equation}
Here   $\mu\in (1,3)$ is the exponent of the underlying power law of the \textit{L\'evy} distribution, and $L \in \mathbb N$ is the truncation size.   These parameters   represent the mobilities of the criminal agents. Different types of agents often assume different mobilities. For example, the parameter $\mu$ for professional criminals is typically lower than that of  amateur criminals.\cite{robbery, individual}  We call the movement pattern defined by (\ref{eq:movq}) and (\ref{eq:weight})  a truncated \textit{L\'evy} flight (TLF). When  $L=\infty$, we call it a \textit{L\'evy} flight. When $L=1$, then (\ref{eq:movq}) and (\ref{eq:weight}) imply 
\begin{equation*}
q_{k \rightarrow i} (t)= \dfrac{A_{i} (t)}{A_{ k-1}(t) + A_{k+1} (t)},\quad \quad i = k-1 \mbox{ or } i=k+1, 
\end{equation*} and we call this a biased random walk (BRW). If the random walk   is unbiased, that is, if $q_{k \rightarrow k-1} (t) = q_{k \rightarrow k-1}= 1/2$, then we call it an unbiased random walk (URW).

%
%
%

%

\textit{Step 3.}
The attractiveness field gets updated  according to the repeat victimization and the broken windows effect.\cite{budd1999burglary, bw2, bw1}  The repeat victimization  is introduced by letting the dynamic attractiveness depend upon previous burglary events at the local site. Whenever there is a burglary event, the local attractiveness will get  increased by an absolute   constant $\theta$. However  it is reasonable to suppose that    this higher probability of burglary at a site has a finite lifetime.  
This increase and decay can be modeled  according to the following update rule 
\begin{equation*}
{B}_k(t+\delta t)={B}_k(t) (1-\omega\delta t)+\theta {E}_k(t), 
\label{eq:B_dis1267}
\end{equation*} 
where   $\omega$ is an absolute constant   representing the decay rate, and  $E_k (t)$ denotes the number of burglary events occurred during the time interval $(t, t+\delta t]$ at site $k$. 
To  further incorporate the broken windows effect, we allow  the dynamic attractiveness field  to  spread spatially from each site to its      nearest neighbours. 
This can be  accomplished by modifying the above equation  as 
%
\begin{equation*}
{B}_k(t+\delta t)=\left[(1-\eta){B}_k(t)+\dfrac{\eta}{2}({B}_{k-1}(t)+{B}_{k+1}(t))\right](1-\omega\delta t)+\theta {E}_k(t),
\label{eq:B_dis25}
\end{equation*}
where $\eta \in (0,1)$ is an absolute constant representing the strength of the   near-repeat victimization effect.
Since on average the attractiveness can be roughly expressed by replacing $E _k(t)$ with $\delta t A _k (t) \dom _k(t)$    according to  (\ref{eq:pr}),    we  finally   set the    evolution of the   dynamic attractiveness term as
\begin{equation}
{B}_k(t+\delta t)=\left[(1-\eta){B}_k(t)+\dfrac{\eta}{2}({B}_{k-1}(t)+{B}_{k+1}(t))\right](1-\omega\delta t)+\theta   \delta t A_k (t) \dom_k (t).
\label{eq:B_dis}
\end{equation} 

\textit{Step 4.}  At each site a new agent is  replaced with rate $\gamma$.

\begin{figure}[htpb]
\centering
\includegraphics[width=0.9\textwidth]{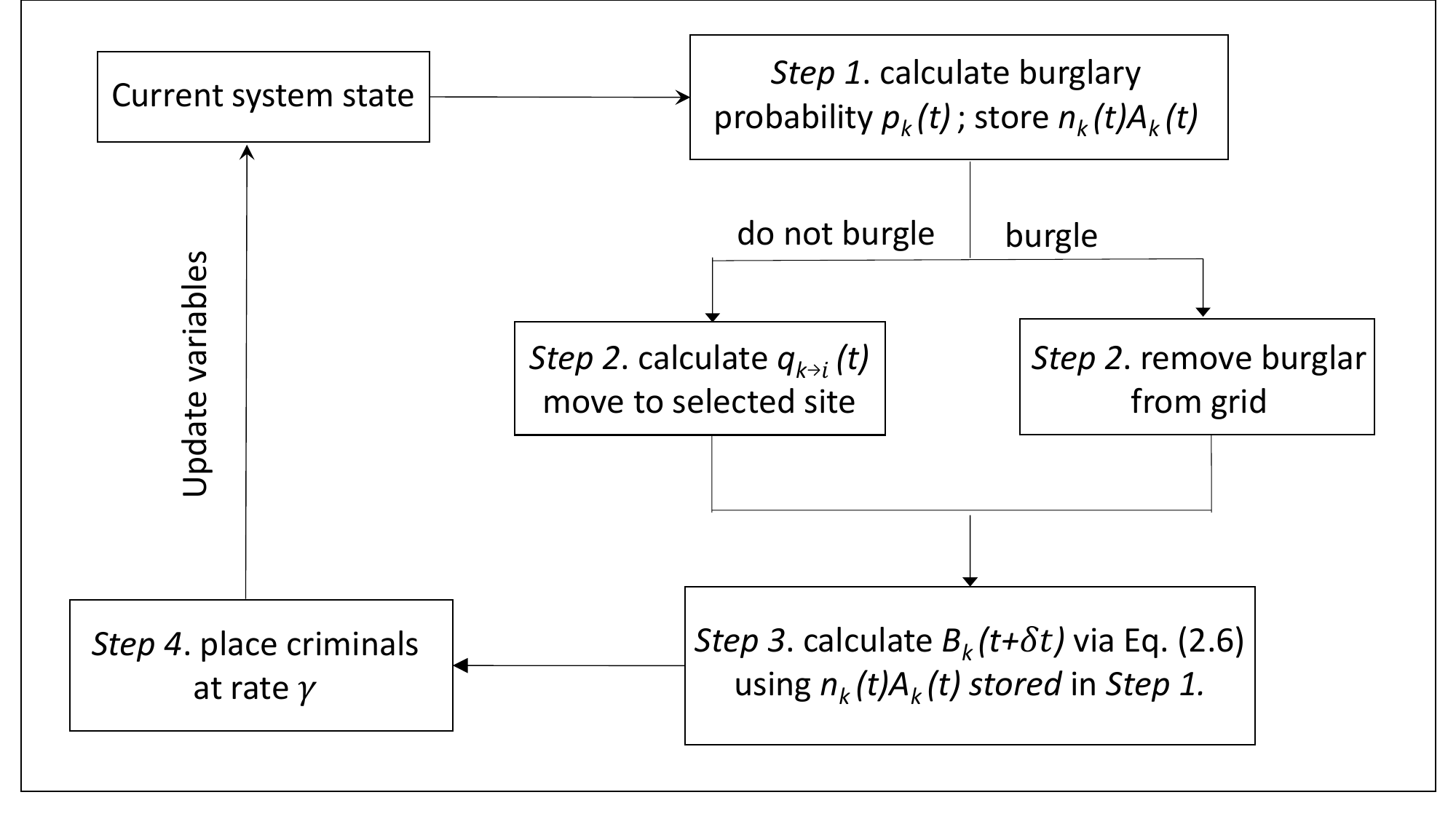}
\caption{Flowchart summarizing the discrete model.}\label{figdiagram1}
\end{figure}

Figure \ref{figdiagram1} presents a visual summary of these four steps in the form of a flow chat. 

To conclude, the discrete mean field equation  of $( A_k (t), n_k (t))$ consists of (\ref{eq:B_dis}) and the following equation
\begin{equation}
\dom_k(t+\delta t) =\sum_{\substack{i \in \mathbb{Z} \\ 1 \leq |i-k| \leq L}}{[1-A_i(t)\delta t]\dom_{i}(t) q_{i \rightarrow k}(t)}+\gamma\delta t. \label{eq:rhoDisc} 
\end{equation}

%

%
%
When $L=1$ and $L=\infty$, 
the assumptions above yield respectively  the random-walk model (RWM) in  Ref. \refcite{uclaModel}, 
and  the  \textit{L\'evy}-flight model (LFM) in Ref.  \refcite{levycrime}. Hence our first task is to see how varying $L$ will affect the  behavior of the  truncated-\textit{L\'evy}-flight model (TLFM). 

\subsection{Computer simulations}
 
\label{sec:compare} 

We simulate   the truncated-\textit{L\'evy}-flight model 
 for several different values of jump length $L$. An example output can be seen in  Fig.      \ref{fig:compA}   below. 
 %
     The  domain is   $[0, 1]$  and ${l}=1/60$.  The computations assume periodic boundary conditions.  Here   the initial conditions (at $t=0$) are taken to be $B_k \equiv 0$ and ${\dom}_k \equiv 1$. The parameters are  $A^0  = 1-0.4\cos(4\pi x)$,
   $\mu = 2.5$, $l = 1/60$, $\delta t = 0.01$,  ${\eta} = 0.1$, $\gamma = 6$, $\omega = 1$,  and $\theta =1$.

\begin{figure}[htpb]
\centering
\begin{subfigure}[b]{0.44\textwidth}
\centering
\includegraphics[width=1  \textwidth]{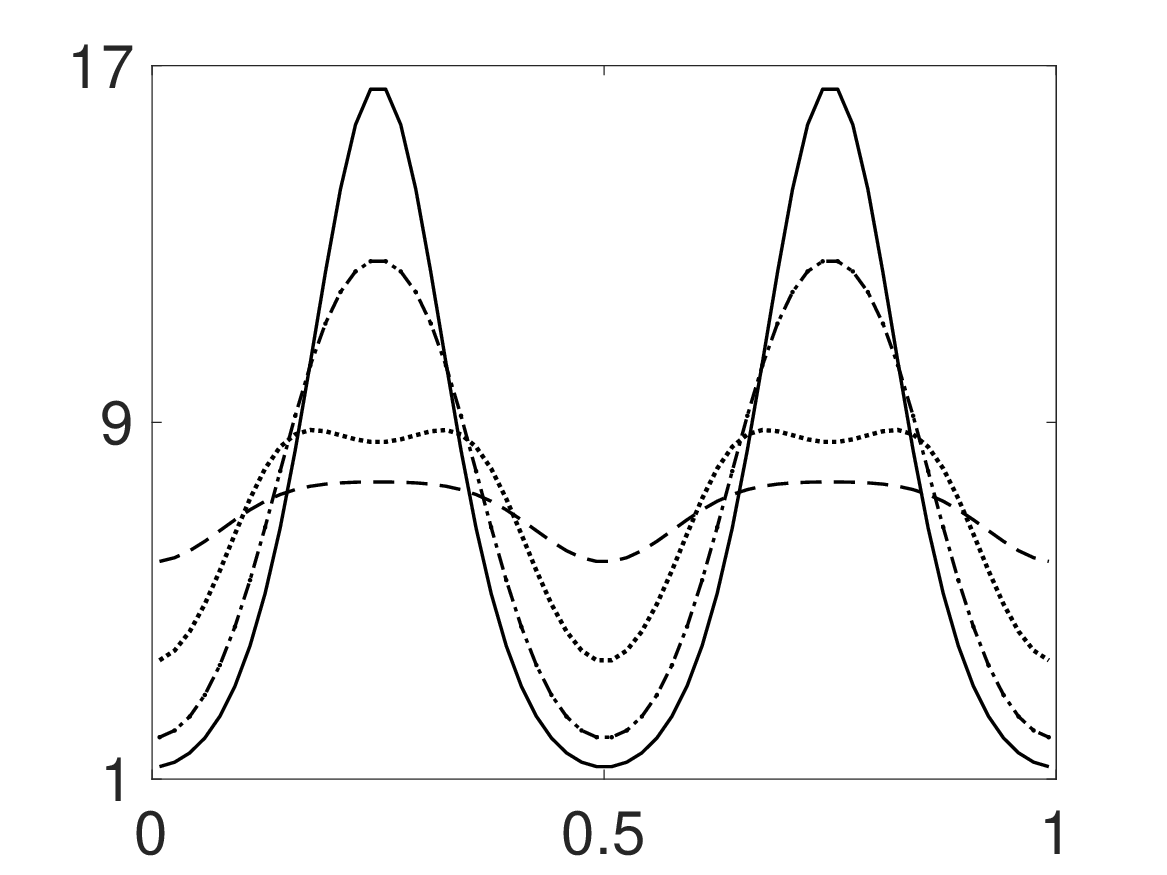}
\caption{$t =8$}
\label{fig:t_10_1}
\end{subfigure}
\hskip -5 mm
\begin{subfigure}[b]{0.585\textwidth}
\centering
\includegraphics[width=1 \textwidth]{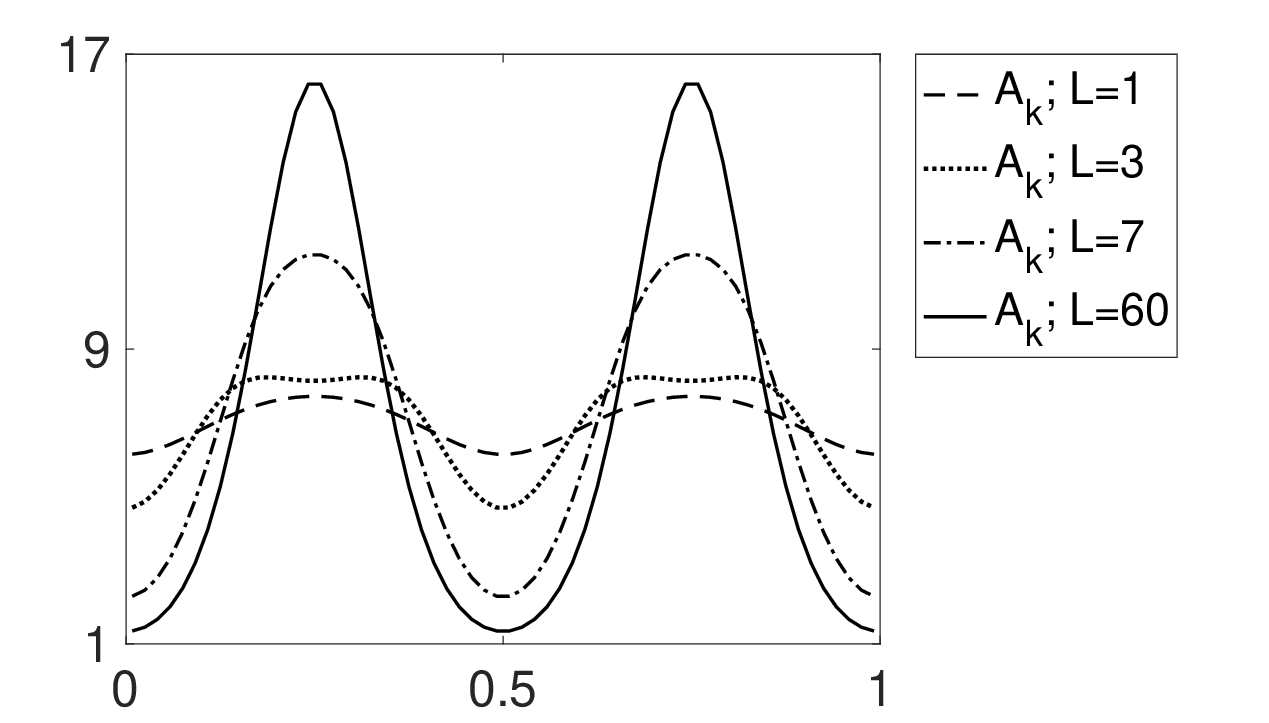}
\caption{$t = 20$}
\label{fig:t_200}
\end{subfigure}
\begin{subfigure}[b]{0.44\textwidth}
\centering
\includegraphics[width=1  \textwidth]{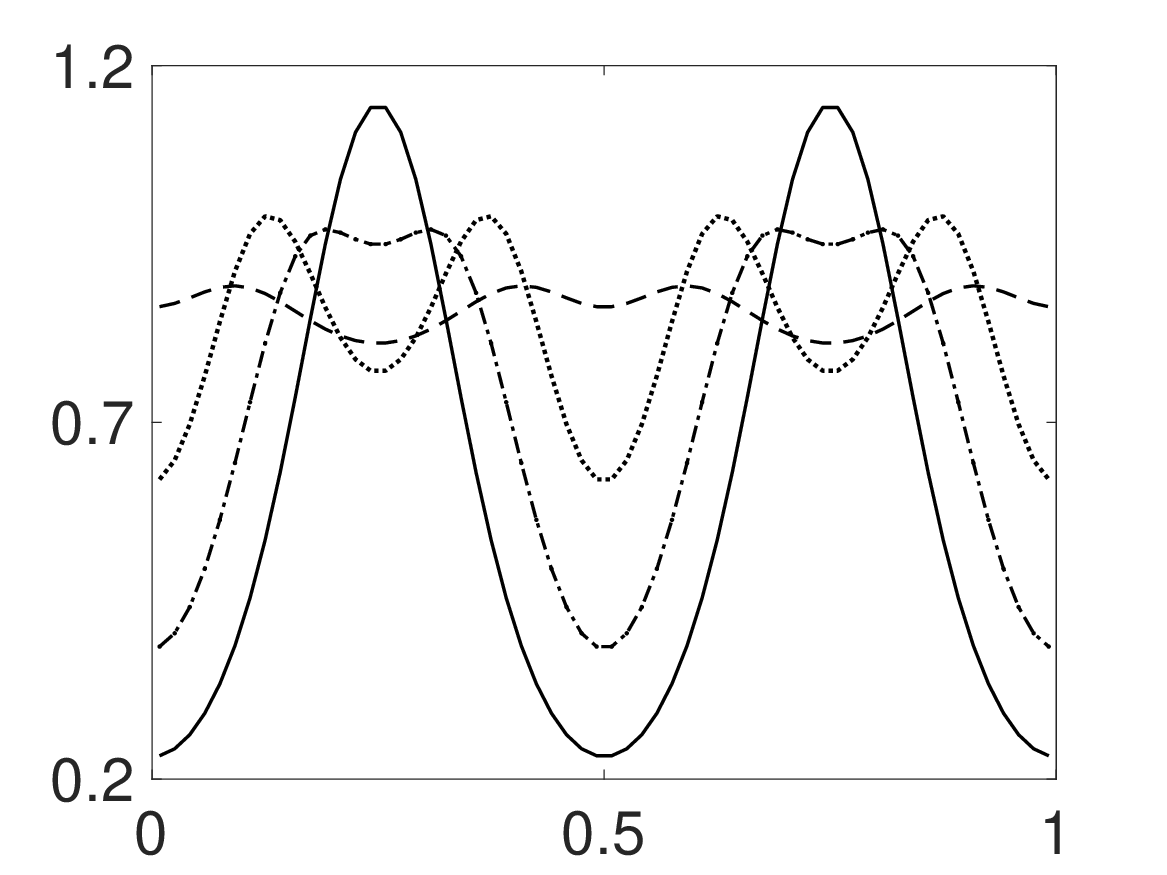}
\caption{$t =8$}
\label{fig:t_10_2}
\end{subfigure}
\hskip -5 mm
\begin{subfigure}[b]{0.585\textwidth}
\centering
\includegraphics[width=1 \textwidth]{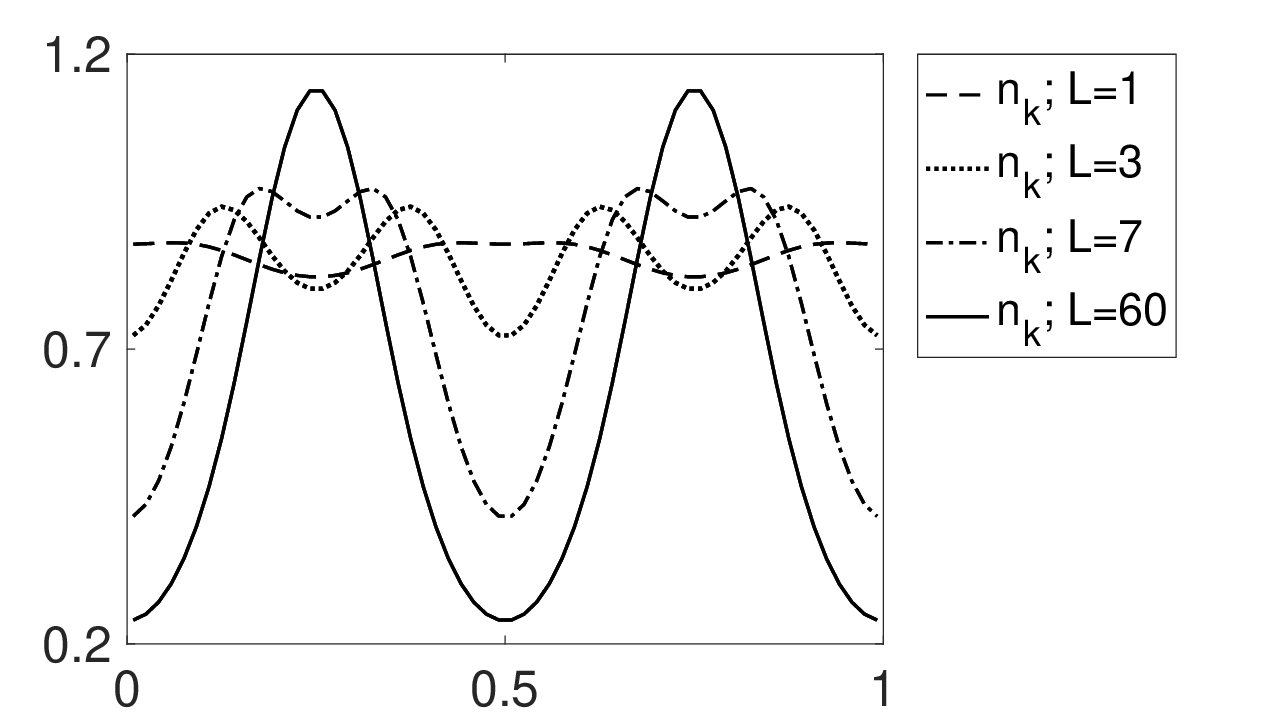}
\caption{$t =20$}
\label{fig:t_10_3}
\end{subfigure}
\caption{Results of  the model (\ref{eq:decomposed})-(\ref{eq:B_dis}) for different values of $L$, using the parameters described in the text. The plots of the attractiveness field are shown in (a), (b), and the  plots of the  criminal number distribution are shown in (c), (d).  For   $L=1$, $3$, $7 $,  and $60$,  they are shown respectively with  dashed lines,     dotted lines,      dash-dotted lines,  and   solid lines.}\label{fig:compA}
\end{figure}
%
%
 We observe that the behavior of the model varies considerably with different choices of  $L$.    This was already noted in the prior work Ref. \refcite{levycrime}. This suggests that more careful analysis should be carried out to connect the two ideas. Here we show that truncation is precisely the correct parameter for this research direction. 
 

\section{Continuum Model}\label{continulimit}

\subsection{Derivation}

In this section, we      derive  an asymptotic theory  when  $\delta t$ and $l$ become small under some suitable spatial-temporal scaling for generic $L\in \mathbb N$. 

We first  derive the dynamics of the continuum version of the attractiveness field.  Following Ref. \refcite{uclaModel},   we observe that the Brownian scaling is a suitable spatial-temporal scaling  for   (\ref{eq:B_dis}). That is, as ${l}$ and $\delta t$ become smaller, the quantity ${l}^2 /\delta t $  remains   constant. 
Using the  same  calculations  as in   Ref. \refcite{uclaModel},  from  (\ref{eq:B_dis}) and (\ref{eq:decomposed}) we infer
\begin{equation}
\dfrac{\partial A}{\partial t}=\dfrac{ l^2\eta}{2 \delta t}A_{xx}-\omega(A-A^0)+ \theta\dom A.
\label{eq:contA}
\end{equation}

The derivation of the dynamics of the continuum version of $\dom_k$, however, is more complicated. 
From (\ref{eq:rhoDisc}) we infer 
\begin{equation}
\dfrac{\dom_k(t+\delta t)-\dom_k(t)}{\delta t} = \dfrac{1}{\delta t}\left[\sum_{\substack{i \in \mathbb{Z} \\ 1 \leq |i-k| \leq L}} \dom   _i(1-A_i\delta t)q_{i \rightarrow k}-\dom   _k\right]+\gamma.
\label{eq:deltat}
\end{equation}
We   define 
\begin{align}
z_{\mu, L}&:=2\sum_{k=1}^{L} \dfrac{1}{k^{\mu}},\label{eq:define_z} \\
\mathcal{L}(f_i)&:=\sum_{\substack{j \in \mathbb{Z} \\ 1 \leq |i-j| \leq L}} \dfrac{f_j-f_i}{(|j-i|l)^{\mu}}.
\label{eq:define_L}
\end{align}
It follows from (\ref{eq:weight}) that
\begin{equation}
\sum_{\substack{i \in \mathbb{Z} \\ 1 \leq |i-k| \leq L}}{w_{i\rightarrow k}}=l^{-\mu}z_{\mu, L}A_i+\mathcal{L}(A_i).
\label{eq:define_weight}
\end{equation}
With (\ref{eq:define_weight}) and (\ref{eq:movq}),  we obtain
\begin{align}
q_{i\rightarrow k}
&=\dfrac{w_{i \rightarrow k}}{l^{-\mu}z_{\mu, L} A_i \left (\dfrac{\mathcal{L}(A_i)}{l^{-\mu}z_{\mu, L} A_i}+1 \right)}  \notag \\
&\sim w_{i \rightarrow k}\left[\dfrac{1}{l^{-\mu}z_{\mu, L} A_i}-\dfrac{\mathcal{L}(A_i)}{(l^{-\mu}z_{\mu, L} A_i)^2}\right] \notag\\
&= \dfrac{A_k}{|i-k|^{\mu}}\left(\dfrac{1}{z_{\mu, L}A_i}-\dfrac{\mathcal{L}(A_{i})l^\mu}{A_{i}^2 z_{\mu, L}^2}\right), \hspace{5mm} 1 \leq |i-k| \leq L.\label{eq:qik} 
\end{align}
 Here   we  have used the fact that     ${1}/({1+x})$ approximates $  1-x$ as long as  $x^2 \sim o (1)$.
Applying (\ref{eq:qik}) to the right-hand-side of (\ref{eq:deltat}), we obtain
\begin{align}
 \!\!\!\!  \!\!\dfrac{\dom_k(t+\delta t)- \dom _k(t)}{\delta t} 
&=\dfrac{1}{\delta t} \!\!\!\!\sum_{\substack{i \in \mathbb{Z} \\ 1 \leq |i-k| \leq L}} \!\!\!\!{ \dom  _i(1-A_i\delta t)   \dfrac{A_k}{|i-k|^\mu}\left(\dfrac{1}{z_{\mu, L} A_i}-\dfrac{\mathcal{L}(A_i)l^\mu}{A_i^2 z_{\mu, L}^2}\right)}-\dfrac{  \dom _k}{\delta t}+\gamma \notag\\
&=\dfrac{A_k}{\delta t} \left[\sum_{\substack{i \in \mathbb{Z} \\ 1 \leq |i-k| \leq L}}(1-A_i\delta t){\dfrac{\dom   _i}{A_i}\dfrac{1}{z_{\mu, L}|i-k|^\mu}}-\dfrac{\dom  _k}{A_k}\right]\notag\\
&\hspace{5mm}-\dfrac{A_k}{\delta t}\sum_{\substack{i \in \mathbb{Z} \\ 1 \leq |i-k| \leq L}} \left[(1-A_i\delta t)\dfrac{\dom_i}{|i-k|^\mu}\dfrac{\mathcal{L}(A_i)l^\mu}{A_i^2 z_{\mu, L}^2}\right]+\gamma. \label{eq:long}
\end{align}
In order to simplify  (\ref{eq:long}),   from  (\ref{eq:define_L}) we infer 
\begin{equation*}
\sum_{{\substack{i \in \mathbb{Z} \\ 1 \leq |i-k| \leq L}}}\!\! \!\!\!\dfrac{\dom  _i}{|i-k|^{\mu}}=\sum_{{\substack{i \in \mathbb{Z} \\ 1 \leq |i-k| \leq L}}} \!\!\!\!\!\! \dfrac{ \dom _i- \dom  _k}{|i-k|^{\mu}}+\sum_{{\substack{i \in \mathbb{Z} \\ 1 \leq |i-k| \leq L}}} \!\!\!\!\!\! \dfrac{\dom   _k}{|i-k|^{\mu}}=l^{\mu}\mathcal{L}( \dom  _k)+z_{\mu, L} \dom  _k \sim z_{\mu, L} \dom  _k,
\label{eq:approx}
\end{equation*}
where   the $O(l^\mu)$ terms have been ignored in the final step.
This together with (\ref{eq:long})  implies that
\begin{align}
\dfrac{\dom  _k(t+\delta t)-\dom  _k(t)}{\delta t}
&=\dfrac{A_k}{\delta t} \!\!\!\!\sum_{\substack{i \in \mathbb{Z} \\ 1 \leq |i-k| \leq L}} \!\!\!\!\!\left[\dfrac{ \dom _i}{A_i} \dfrac{1}{z_{\mu, L}|i-k|^{\mu}}-\delta t\dfrac{ \dom  _i}{|i-k|^{\mu}z_{\mu, L}}-\dfrac{\dom_k}{A_k}\dfrac{1}{z_{\mu, L}|i-k|^{\mu}}\right] \notag \\
&\hspace{5mm}-\dfrac{A_k}{\delta t} \!\!\!\! \sum_{\substack{i \in \mathbb{Z} \\ 1 \leq |i-k| \leq L}}\left[\dfrac{\dom_i}{|i-k|^\mu} \!\!\!\!\!\ \dfrac{\mathcal{L}(A_i)l^\mu}{A_i^2z_{\mu, L}^2}-\dfrac{\dom_i\mathcal{L}A_i}{A_i zz_{\mu, L}^2|i-k|^{\mu}}l^{\mu}\delta t \right]+\gamma\notag \\
&\sim\dfrac{A_k}{\delta t}\!\!\!\!\! \!\!\!\sum_{\substack{i \in \mathbb{Z} \\ 1 \leq |i-k| \leq L}} \!\!   \left[{\dfrac{\dfrac{\dom_i}{A_i}\! -\!\dfrac{\dom_k}{A_k}}{|i-k|^\mu z_{\mu, L}}}\!-\!\dfrac{\dom_i}{|i-k|^\mu}\dfrac{\mathcal{L}(A_i)l^\mu}{A_i^2z_{\mu, L}^2} \!-\! \delta t\dfrac{\dom_i}{|i-k|^\mu z_{\mu, L}}\right] \!\! +\!\!\gamma\notag \\ 
&\sim\dfrac{l^\mu}{z_{\mu, L}\delta t}\left[A_k\mathcal{L}\left(\dfrac{\dom_k}{A_k}\right)-\dfrac{\dom_k\mathcal{L}(A_k)}{A_k}\right]- A_k \dom_k+\gamma. \label{eq:short}
\end{align}
Here  at the second step, all the $O(l^{\mu}\delta t)$ terms have been ignored in the summation.
We now simplify  
\begin{equation}
\mathcal{L}(A_k)=\sum_{\substack{j \in \mathbb{Z} \\ 1 \leq |j-k| \leq L}} {\dfrac{A_j-A_k}{(|j-k|l)^\mu}}. 
\label{eq:riemann}
\end{equation}
%
%
Let $x=kl$ and then $A_k = A(x)$. When  $l$ is small,  we can  apply the Taylor expansion  to the integrand   near $x$   and obtain
\begin{align}
\mathcal{L}(A_k)&=\!\!\!\!\!\sum_{\substack{j \in \mathbb{Z} \\ 1 \leq |j-k| \leq L}}{{\!\!\!\!\! (|j-k|l)^{-\mu}}  \left[{A_x(kl)(j-k)l+A_{xx}(kl)\dfrac{((j-k)l)^2}{2}+\!O((|j-k|l)^3)}\right]} \notag\\
&\sim  \left[\sum_{\substack{j \in \mathbb{Z} \\ 1 \leq |j-k| \leq L}}{\dfrac{A_x(kl) (j-k)l}{ (|j-k|l) ^\mu} }+  \sum_{\substack{j \in \mathbb{Z} \\ 1 \leq |j-k| \leq L}}{\dfrac{A_{xx}(kl)( (j-k)l) ^2}{2 (|j-k|l) ^\mu} }\right] \notag\\
&=   \frac{1}{2}\sum_{\substack{j \in \mathbb{Z} \\ 1 \leq |j-k| \leq L}}{{(|j-k|l)^{2-\mu}} A_{xx} (kl)   } \notag\\
&= {l^{2-\mu}}\sum_{j=1} ^{L}j^{2-\mu}A_{xx}(kl) .
\label{eq:longer1}
\end{align}
Here at the second step, the $O((|j-k|l)^{3-\mu})$ terms and lower order terms are all ignored as $\mu < 3$.
 We then obtain
\begin{equation}\label{eq:longer}
\mathcal{L}(A_k) ={l^{2-\mu}}z ^\ast_{\mu, L} A_{xx}(kl)  ,
\end{equation}
where
\begin{equation}
  z ^\ast_{\mu, L}:= \sum_{j=1} ^{L}j^{2-\mu}. 
\end{equation}
From (\ref{eq:longer}) and  (\ref{eq:short}) we infer
\begin{equation}
\dfrac{\partial \dom}{\partial t}= \mathcal{D}  \vec{\nabla} \cdot \left [ \vec{\nabla} \dom - \dfrac{2 \dom}{A} \vec{\nabla}A \right]-A\dom+\gamma, 
\label{eq:contrho}
\end{equation} 
where 
\begin{equation}\label{eq:D1}
\mathcal{D} =\dfrac{l^2  }{\delta t }\frac{z ^\ast_{\mu, L}}{z_{\mu, L}} .
\end{equation}
Here $\mathcal D$ is the diffusion coefficient which depends on $\mu$ and $L$. Particularly when $L=1$, then $\mathcal D=l^2/2 \delta t$.

To validate the continuum model
 we next perform direct numerical simulations   and compare  it with the discrete model. 

\begin{remark}\label{rmk4}
When $L=1$ and  $L=\infty$,  we recall that  the mean field continuum equations of 
 the random-walk model (RWM)   and the  \textit{L\'evy}-flight model  (LFM)
have been derived 
     in Refs.  \refcite{uclaModel} and  \refcite{levycrime}:
\begin{equation}\label{eq:walk}\displaystyle
\hspace{-34 mm}\mbox{Continuum RWM }
 \begin{cases}
& \!\!\!\!\!\dfrac{\partial A}{\partial t}=\dfrac{l^2\eta}{2 \delta t} A_{xx}-\omega (A-A^0)+ \theta A{\dom},\\
\\
& \!\!\!\!\!\dfrac{\partial \dom}{\partial t}=\dfrac{l^2}{2 \delta t} \vec{\nabla} \cdot \left [ \vec{\nabla} \dom - \dfrac{2 \dom}{A} \vec{\nabla}A \right]    -{A}{{\dom}}+\gamma, 
\end{cases}
\end{equation}
\begin{equation}\label{eq:flight}
 \displaystyle\mbox{Continuum LFM }
\begin{cases}
&\!\!\!\!\!\dfrac{\partial A}{\partial t}=\dfrac{l^2\eta}{2 \delta t}   A_{xx}-\omega (A-A^0)+ \theta A{\dom},\\
\\ 
& \!\!\!\!\!\dfrac{\partial \dom}{\partial t} = \dfrac{l^{2s}}{\delta t}\dfrac{\sqrt{\pi}2^{-2s}|\Gamma(-s)|}{z\Gamma(s+\dfrac{1}{2})  }\left[{A}\Delta^s (\dfrac{{{\dom}}}{{A}})-\dfrac{{{\dom}}}{{A}}\Delta^s {A}\right]-{A}{{\dom}}+\gamma. 
\end{cases}
\end{equation}
Here
$  s=\left ({\mu-1}\right)/{2}$, $z = 2 \sum_{k=1}^\infty k ^{-\mu}$,  and $\Gamma(\cdot)$ denotes  the gamma function. We also note that   when $L=1$,  (\ref{eq:contrho}) coincides with (\ref{eq:walk})$_2$ as desired. 
%
%
For generic $L\in \mathbb N$,   however,   (\ref{eq:walk}) and (\ref{eq:flight}) may not be applicable anymore seen from   Fig. \ref{fig:compA},  and this is why we need to    derive  new continuum  equations for the  truncated-\textit{L\'evy}-flight model.

Furthermore, in (\ref{eq:contrho})  the Laplacian operator replaces the   fractional Laplacian operator  in (\ref{eq:flight}).  This happens essentially because the  
     infinitesimal generator of  truncated \textit{L\'evy} flights is a  local operator. 
  An analogous fact is that the  independent sum of  $N$ truncated \textit{L\'evy} flights can be approximated by a Gaussian process when $n$ is large. \cite{stochastic}  
\end{remark}

\subsection{Computer simulations}\label{sec:comparisons}

Figs. \ref{fig:levy_plot}-\ref{fig:mu_plot} below show the comparison between  the discrete and the continuum truncated-\textit{L\'evy}-fight models. The computation overall assumes  periodic boundary conditions.  The   algorithm used for the continuum simulation is very similar to  the one applied to  the continuum  random-walk model   (see   (3.11)-(3.13)  in Ref.   \refcite{uclaModel}).
   Particularly,  we use a semi-implicit time discretization as follows:
\begin{align}\displaystyle
A^{(m+1)} &=A^{(m)} + \Delta t \left (\eta A^{(m)}_{xx}-A^{(m)}+A^{(m)} n^{(m)}+A^0\right ) , \label{eq:num_tlf1}\\
n^{(m+1)} &=n^{(m)} + \mathcal {D}\Delta t \left[n^{(m)}_{xx}-\left (\dfrac{2n A_x^{(m+1)}}{A^{(m+1)}} \right )_x\right]+\Delta t(-A^{(m+1)}n^{(m)}+\gamma). 
\label{eq:num_tlf}
\end{align} Here  $f^{(m)}$ represents a quantity $f$ at $m$th time step.


 %
 %
 %
In Fig.   \ref{fig:levy_plot},  we set $L$ as $1/l$. We  include the continuum   \textit{L\'evy}-flight model (\ref{eq:flight}) with the equivalent parameters   in the comparison, as  an implicit jump range of $L=1/l$ was used   in  
the  discrete simulation of the  \textit{L\'evy}-flight model.   \cite{levycrime}
Fig. \ref{fig:L_plot}   displays  the comparison  of the    discrete and the continuum models   for several different values of $L$, and Fig. \ref{fig:mu_plot}    displays the  comparison  for different values of $\mu$. 

In all  these cases,  we observe  a  good   agreement   between the discrete  and  the continuum models  all the way to the boundary.   In Figs. \ref{fig:levy_plot}(a) and \ref{fig:levy_plot}(b),  the continuum truncated-\textit{L\'evy}-flight model
fits better than  the continuum \textit{L\'evy}-flight model with the discrete model.

\begin{figure}[htpb]
\centering
\hskip -1.8 mm
\begin{subfigure}[b]{0.44\textwidth}
\centering
\includegraphics[width=1.0  \textwidth]{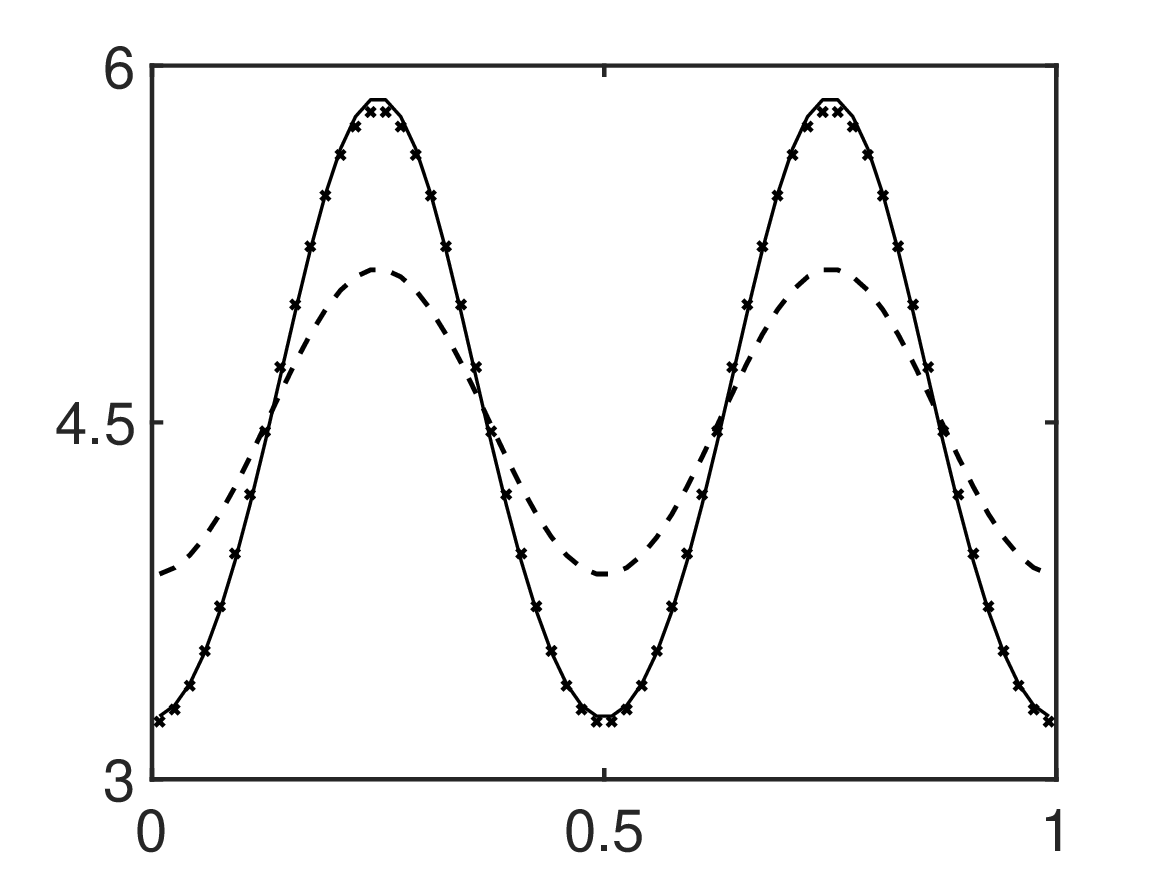}
\caption{$t = 6$}
\label{fig:t_6_1}
\end{subfigure}
\hskip -4.5 mm
\begin{subfigure}[b]{0.59\textwidth}
\centering
\includegraphics[width=1   \textwidth ]{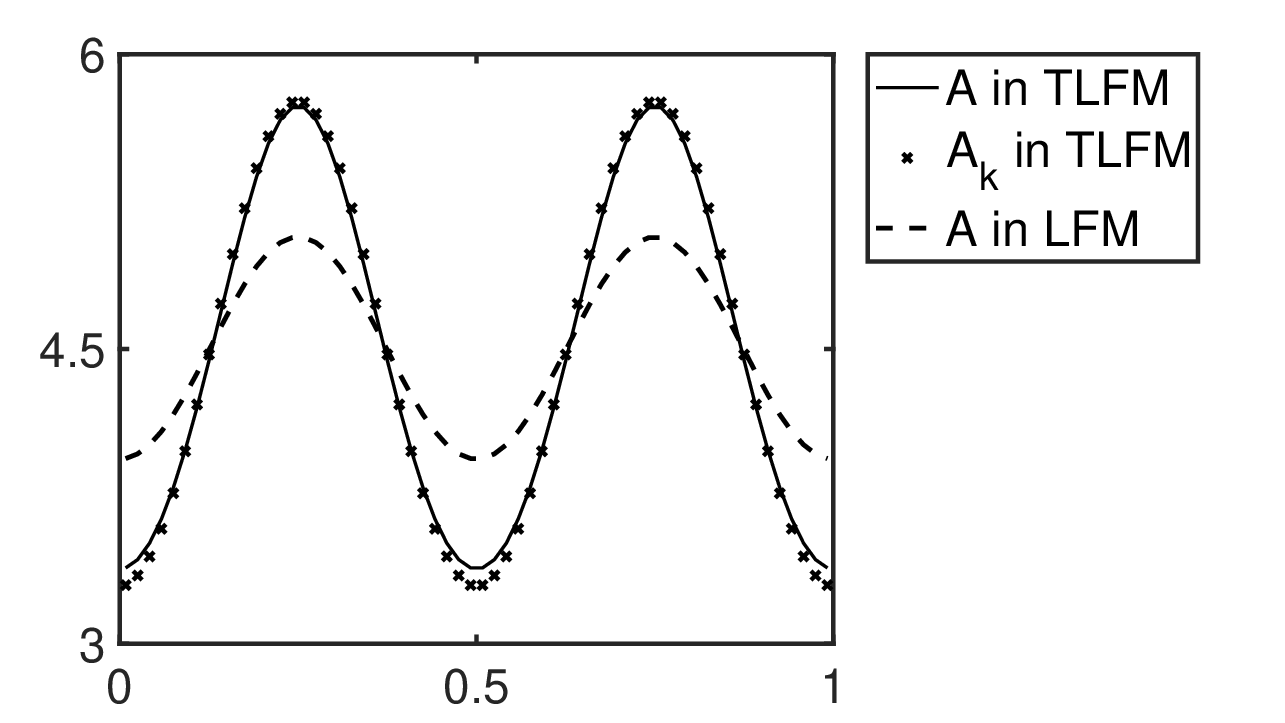}
\caption{$t = 12$}
\label{fig:t_12_1}
\end{subfigure}
\begin{subfigure}[b]{0.44\textwidth}\
\centering
\includegraphics[width=1  \textwidth]{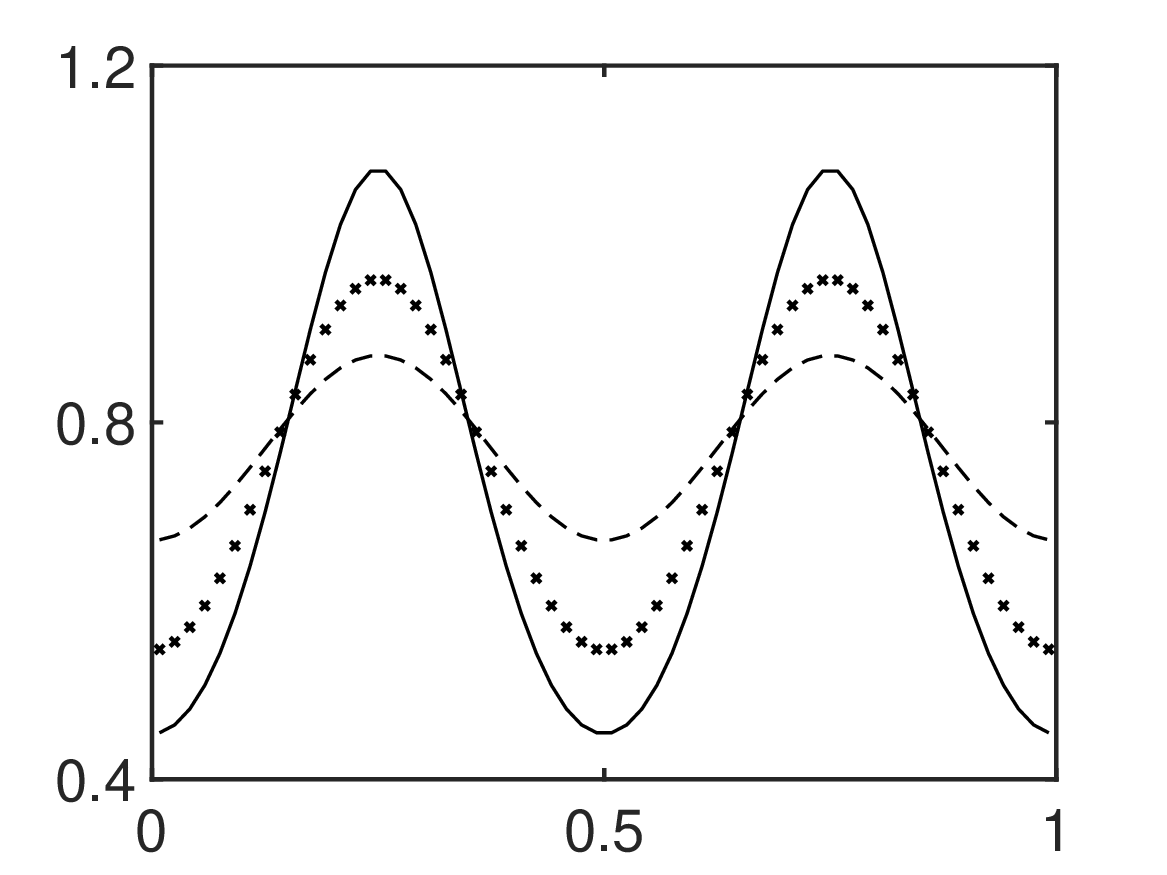}
\caption{$t = 6$}
\label{fig:t_6_2}
\end{subfigure}\hskip -4.5 mm
\begin{subfigure}[b]{0.59\textwidth}
\centering
\includegraphics[width=1  \textwidth ]{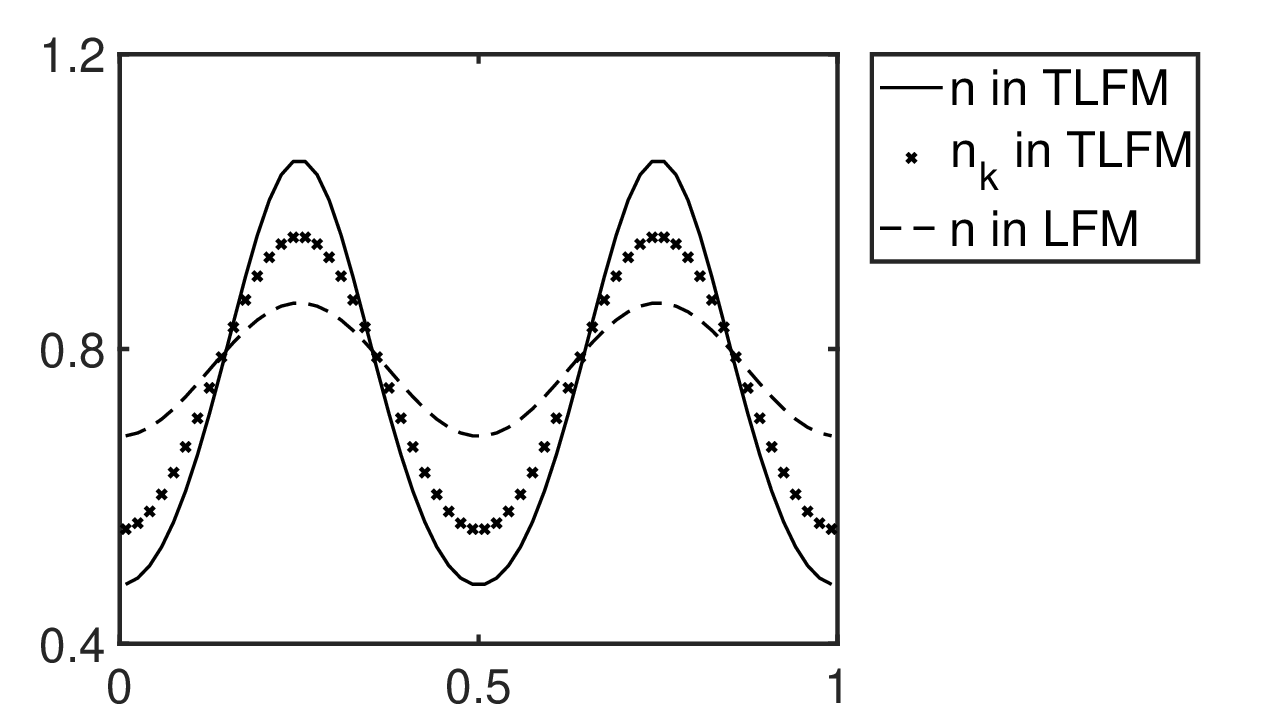}
\caption{$t = 12$}
\label{fig:t_12_2}
\end{subfigure}
\caption{Comparisons of the discrete and the continuum truncated-\textit{L\'evy}-flight models, and the continuum \textit{L\'evy}-flight model  with the equivalent parameters. The plots of the attractiveness field are shown in (a), (b), and those of the criminal number distribution are shown in (c), (d).   
The discrete model   (\ref{eq:decomposed})-(\ref{eq:B_dis}) is shown with 
 cross lines,   the continuum truncated-\textit{L\'evy}-flight model   (\ref{eq:contA}), (\ref{eq:contrho}) is shown with solid lines, and the continuum \textit{L\'evy}-flight model (\ref{eq:flight}) is shown with dashed lines.    
 Here  $L =1/l=60$, ${\eta} = 0.55$, $\gamma = 3.5$,  and all the other parameters and data are the same as in Fig. \ref{fig:compA}.   The system enters a steady state   at roughly  $t=12$.
}
\label{fig:levy_plot}
\end{figure}
%
\begin{figure}[htpb]
\centering
\begin{subfigure}[b]{0.355   \textwidth}
\centering
\includegraphics[width=1.0\textwidth]{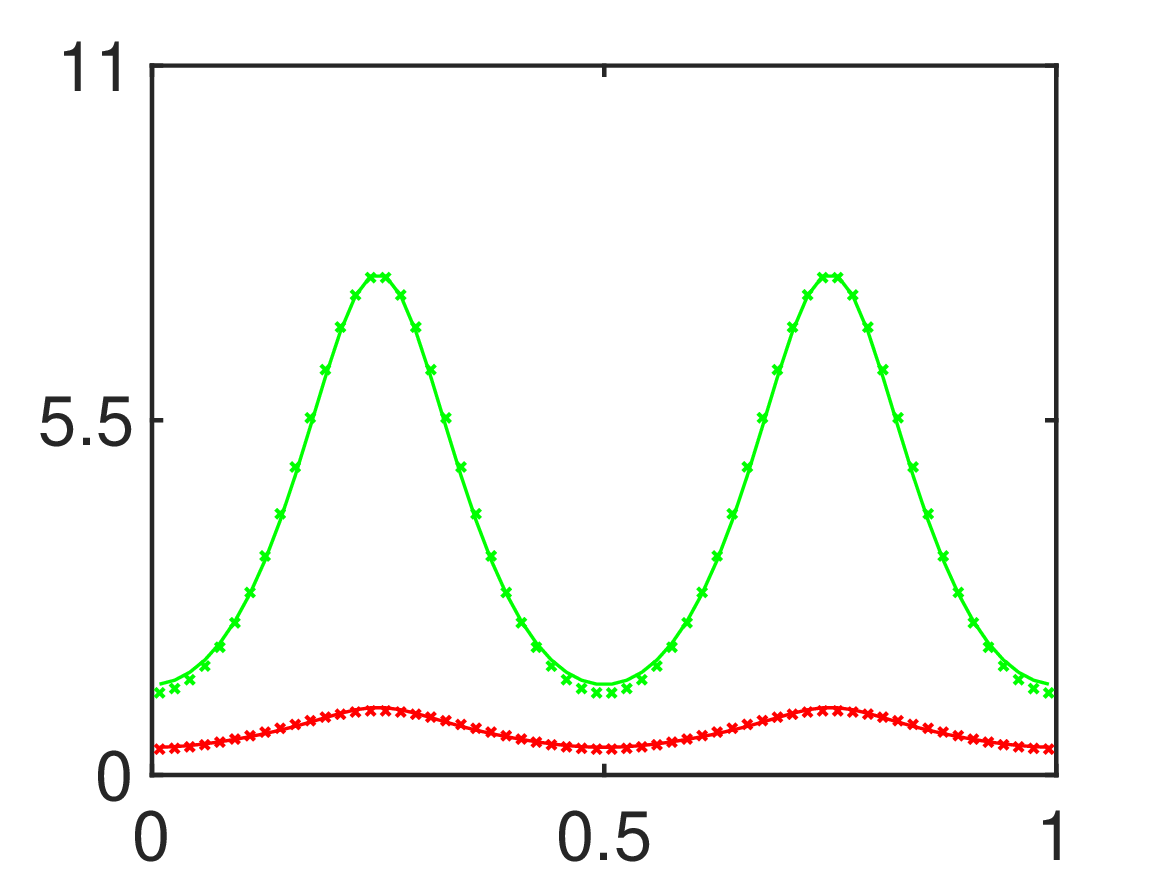}
\caption{$L =3$}
\label{fig:L_3}
\end{subfigure} 
\hskip -5.5  mm
\begin{subfigure}[b]{0.355  \textwidth}
\centering
\includegraphics[width=1.0 \textwidth]{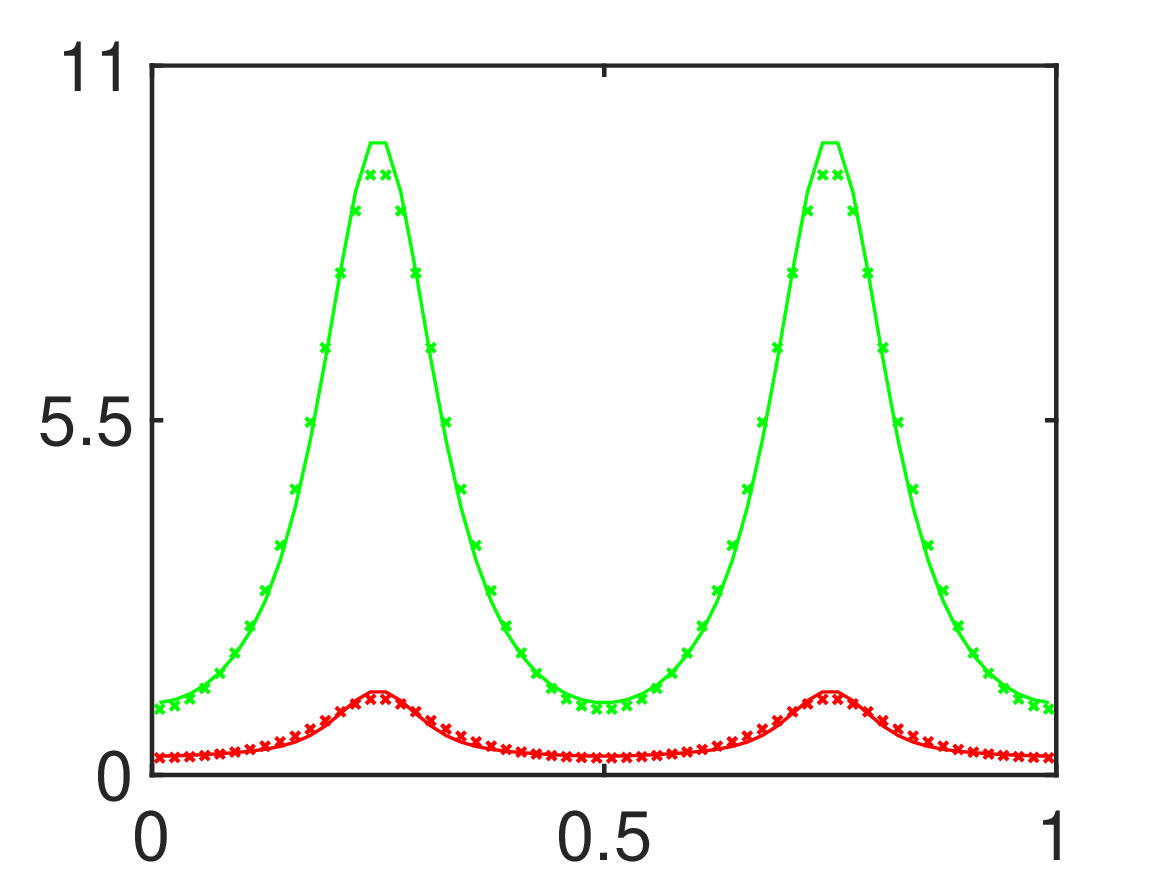}
\caption{$L = 7$}
\label{fig:L_18}
\end{subfigure} 
\hskip -5.5  mm
\begin{subfigure}[b]{0.355  \textwidth}
\centering
\includegraphics[width=1.0 \textwidth]{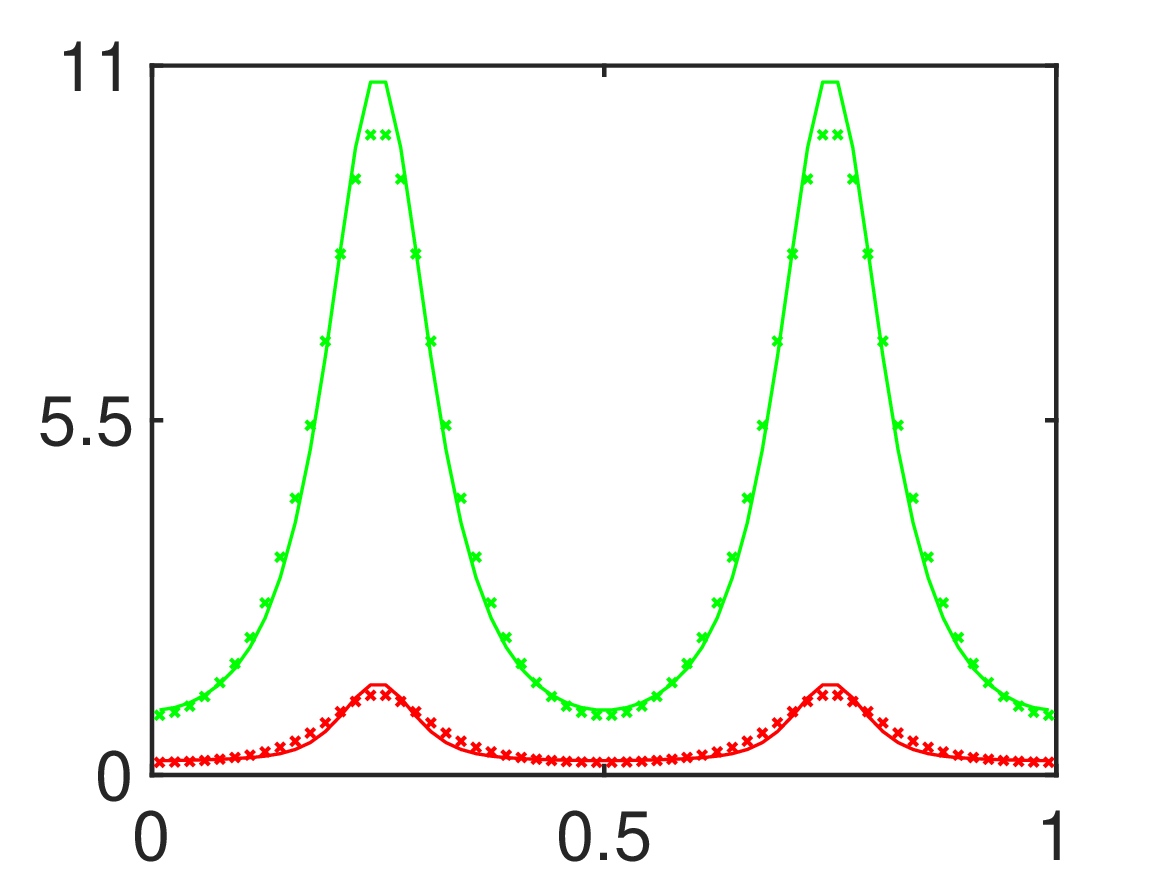}
\caption{$L = 12$}
\label{fig:L_27}
\end{subfigure}
\caption{Comparisons of the discrete  and the continuum truncated-\textit{L\'evy}-flight models    for  different values of $L$. The shots are taken at $t = 2$.  The discrete model (\ref{eq:decomposed})-(\ref{eq:B_dis}) is shown with the cross lines, and  the continuum model  (\ref{eq:contA}) and (\ref{eq:contrho})  is shown with the solid lines. 
The plots of the attractiveness field   are shown with the green  lines,  and those of the criminal number distribution    are shown with the red   lines.   Here 
all the   parameters and data are the same as in Fig.  \ref{fig:compA} except for $\eta = 0.12$ and $L$ as indicated. 
  }
\label{fig:L_plot}
\end{figure}
\vspace*{10pt}
\begin{figure}[htpb]
\centering
\begin{subfigure}[b]{0.355  \textwidth}
\centering
\includegraphics[width=1.0  \textwidth]{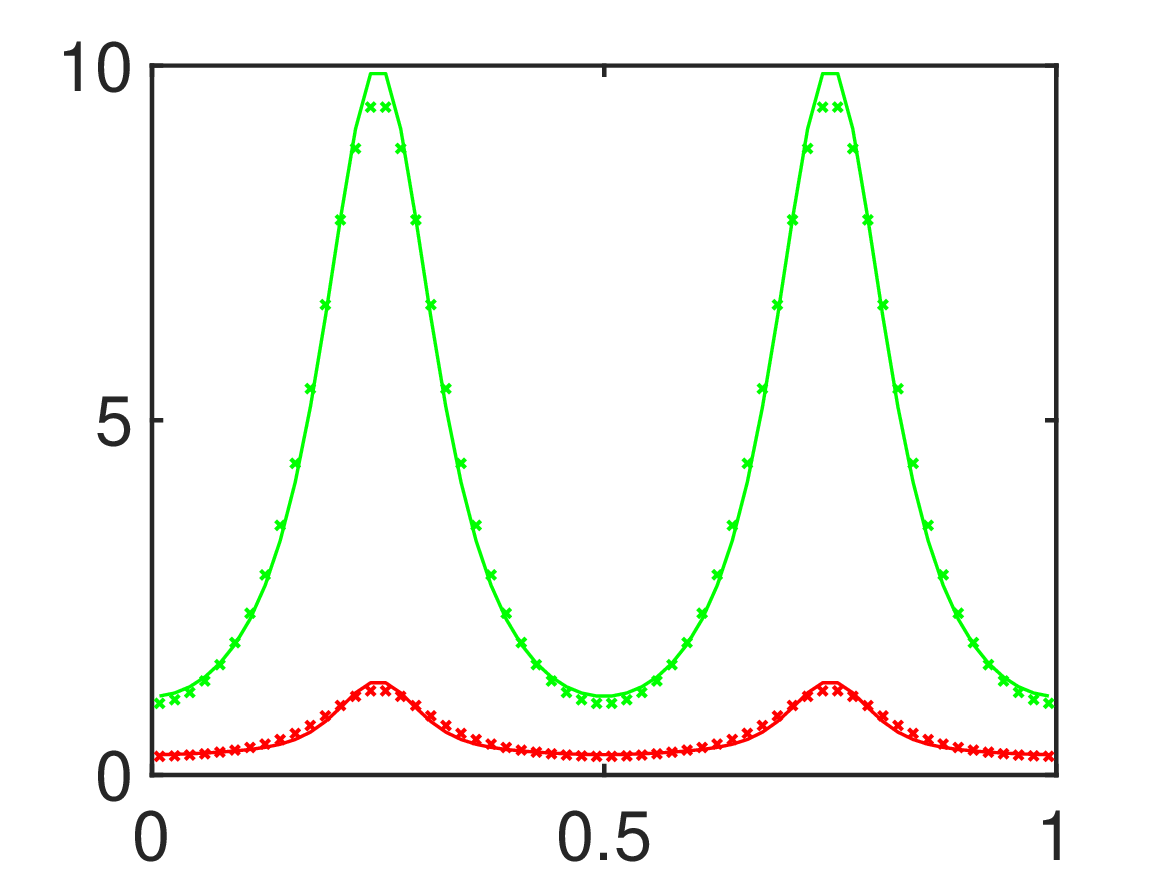}
\caption{$\mu = 1.1$}
\label{fig:mu_1_5}
\end{subfigure}
\hskip -5.5    mm
\begin{subfigure}[b]{0.355  \textwidth}
\centering
\includegraphics[width=1.0  \textwidth]{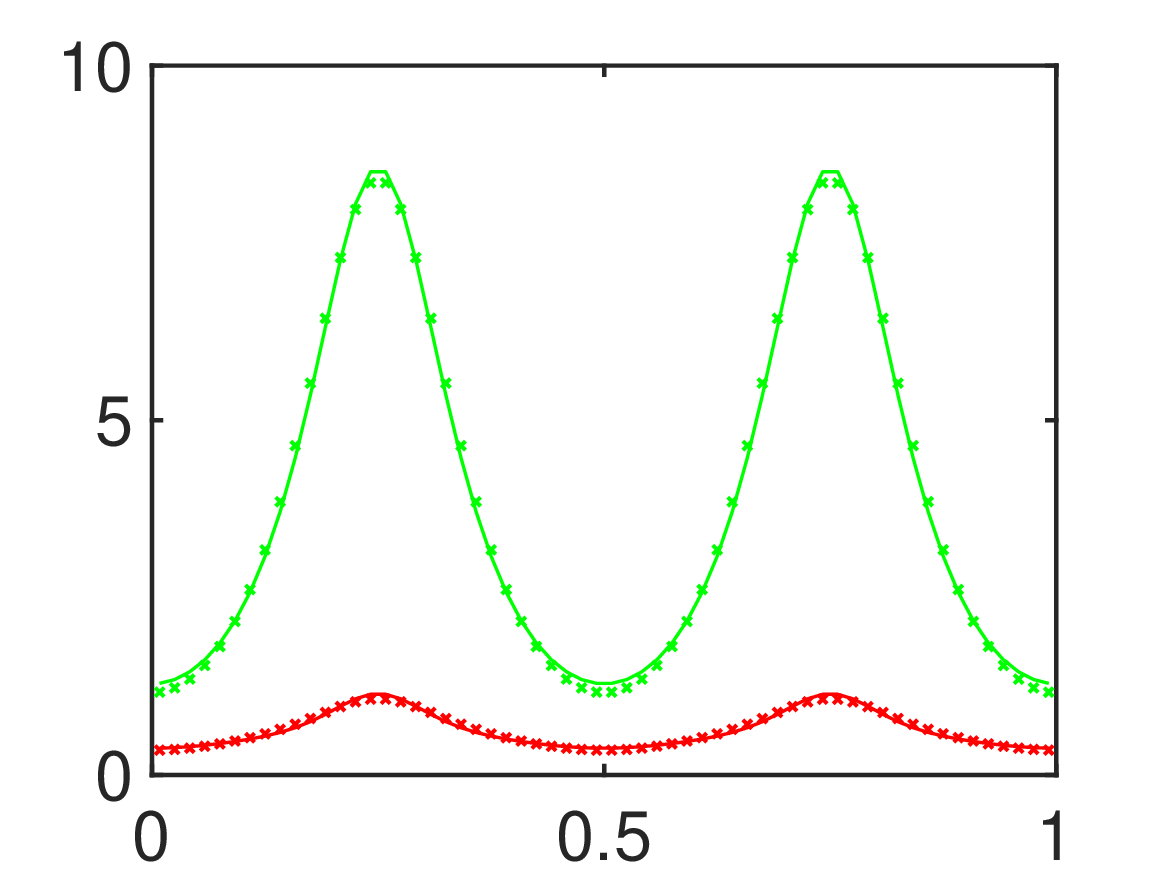}
\caption{$\mu = 2$}
\label{fig:mu_2}
\end{subfigure}
\hskip -5.5   mm
\begin{subfigure}[b]{0.355  \textwidth}
\centering
\includegraphics[width=1.0  \textwidth]{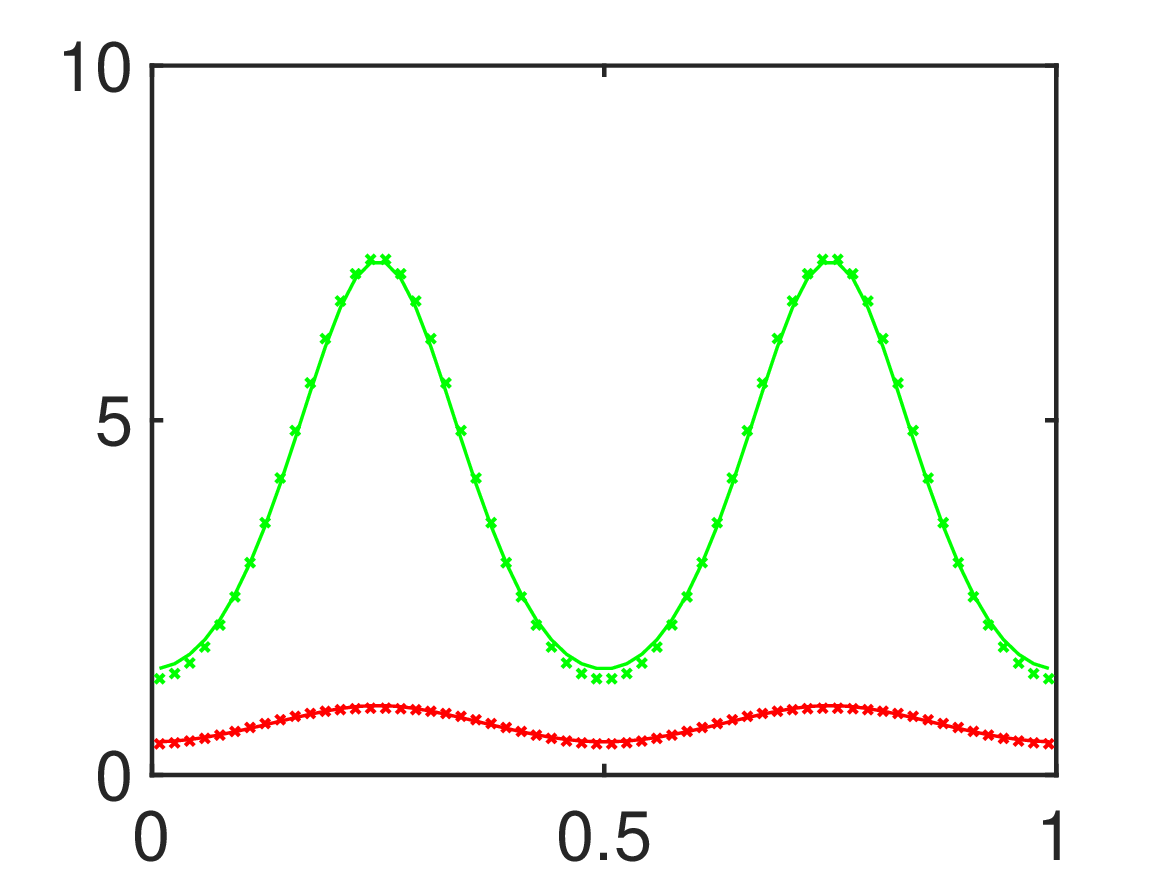}
\caption{$\mu = 2.9$}
\label{fig:mu_2_5}
\end{subfigure}
\caption{Comparisons of the  discrete and the continuum truncated-\textit{L\'evy}-flight models for     different values of $\mu$. The shots are taken at $t = 2$.   The discrete model (\ref{eq:decomposed})-(\ref{eq:B_dis}) is shown with the cross lines, and  the continuum model  (\ref{eq:contA}) and (\ref{eq:contrho})  is shown with the solid lines. 
The plots of the attractiveness field   are shown with the green  lines,  and  those of the criminal number distribution    are shown with the red   lines.
  Here  all the   parameters and data are the same as in Fig.  \ref{fig:compA} except for $\eta = 0.12$, $L=3$,  and  $\mu$ as indicated. 
}
\label{fig:mu_plot}
\end{figure}

%

\subsection{Linear stability analysis}\label{se:linear}

In this section,  we analyze the formation  of the  hotspots (spatial-temporal collections of criminal activities) as observed
in the previous simulations and develop a stability condition. Since  the continuum equations (\ref{eq:contA}) and (\ref{eq:contrho}) are very similar to (\ref{eq:walk})$_1$ and (\ref{eq:walk})$_2$,  except for a modified diffusion coefficient, 
   the previous stability analysis for the random-walk model  (see e.g. (3.21) in Ref. \refcite{uclaModel}) can be extended directly to suit for the truncated-\textit{L\'evy}-flight model.

  As in Refs. \refcite{uclaModel} and \refcite{levycrime}, we first  rescale the variables in the continuum equations:
\begin{equation}\label{eq:repa}
A={A}^\ast \omega,   \quad \dom = \dfrac{{\dom}^\ast \omega}{\theta}, \quad 
t=\dfrac{{t}^\ast}{\omega}, \quad {\eta}^\ast=\dfrac{l^2\eta}{2\omega\delta t}. 
\end{equation}
Applying      (\ref{eq:repa})
to (\ref{eq:contA}) and (\ref{eq:contrho}), we obtain (the $\ast$'s are omitted)
\begin{align}\displaystyle
\label{eq:A_con_bL1}
  \dfrac{\partial A}{\partial t}&=\eta A_{xx}-A+\alpha+A{\dom},\\
\label{eq:rho_con_bL1}
\dfrac{\partial \dom}{\partial t}&=\bar{ \mathcal{ D} } \vec{\nabla} \cdot \left [ \vec{\nabla} \dom - \dfrac{2 \dom}{A} \vec{\nabla}A \right]-A{\dom}+\beta,
\end{align}
where 
\begin{equation}
\bar{ \mathcal{ D} }   =\dfrac{l^2  }{\delta t }\frac{z ^\ast_{\mu, L}}{\omega z_{\mu, L}}, \hspace{5mm} \alpha=\dfrac{A^0}{\omega},\hspace{5mm} \beta=\dfrac{\gamma\theta}{\omega^2}. 
\label{eq:D_1}
\end{equation} 
 Let the steady states be $(\bar{A}, \bar{{\dom}})$, 
\begin{align}
\label{steady}
\bar{A} = \alpha + \beta, \qquad \bar{{\dom}}=\dfrac{\beta}{\alpha+\beta}. 
\end{align}
We find the following      stability conditions of the system around the homogeneous steady states:
\begin{theorem}\label{thm:stability}
When $\bar{{\dom}}<1/3$,  the homogeneous equilibrium in (\ref{steady}) is stable. When $\bar{{\dom}} > 1/3$,  the equilibrium is unstable if  
%
\begin{equation}
\eta < \bar{ \mathcal{D}}  \dfrac{ 3\bar{{\dom}}+1- \sqrt{12 \bar {{\dom}}}}{\bar{A}}.
\label{eq:unstable}
\end{equation}
\end{theorem}
\begin{proof} The proof is very similar to  that  in  Ref.  \refcite{uclaModel}, that is, we  apply a linear Turing stability analysis on (\ref{eq:A_con_bL1}) and (\ref{eq:rho_con_bL1}) around the homogeneous steady state. 
We   decompose  the solutions as perturbations from the steady states:
\begin{align}
\label{eq:turb}
A(x,t) = \bar{A}+\delta_A e^{\sigma t}e^{ikx},\quad {\dom}(x,t) = \bar{{\dom}}+\delta_{\dom} e^{\sigma t}e^{ikx}.
\end{align}
Substituting (\ref{eq:turb}) into (\ref{eq:A_con_bL1}) and (\ref{eq:rho_con_bL1}), we obtain
\begin{align}
\label{matrixeq}
\begin{bmatrix}
-\eta|k|^2 - 1 + \bar{{\dom}} & \bar{A}\\
\dfrac{2\bar{{\dom}}}{\bar{A}}\bar   \bar{ \mathcal{ D} }    |k|^2 - \bar{{\dom}} & -  \bar{ \mathcal{ D} }  |k|^2 - \bar{A}
\end{bmatrix}
\begin{bmatrix}
\delta_A\\
\delta_{\dom}
\end{bmatrix}
=\sigma
\begin{bmatrix}
\delta_A\\
\delta_{\dom}
\end{bmatrix}. 
\end{align}
We solve for the eigenvalue problem (\ref{matrixeq}).  We first rewrite it  as
\begin{align}
\label{eq:matrix3}
\begin{bmatrix}
-\eta|k|^2 - 1 + \bar{{\dom}}-\sigma & \bar{A}\\
\dfrac{2\bar{{\dom}}}{\bar{A}} \bar{ \mathcal{ D} }  |k|^2 - \bar{{\dom}} & -  \bar{ \mathcal{ D} }  |k|^2 - \bar{A}-\sigma
\end{bmatrix}
\begin{bmatrix}
\delta_A\\
\delta_{\dom}
\end{bmatrix}
=0.
\end{align}
Setting the determinant of the square matrix on the left-hand-side as zero, we obtain
\begin{equation}
\sigma^2 - \tau \sigma + \delta=0,
\label{eq:quadratic2}
\end{equation}
where
\begin{align}
\tau &= -\bar{ \mathcal{ D} }    |k|^2 - \eta |k|^2 - \bar{A} - 1 + \bar{{\dom}},\\ 
\delta &=\bar{ \mathcal{ D} }   |k|^2(\eta |k|^2 + 1 - 3\bar{{\dom}})+ \eta |k|^2 \bar{A}+ \bar{A}.
\label{eq:delta}
\end{align}
The equilibrium is stable if   both solutions to (\ref{eq:quadratic2}) have negative real parts. Since $\alpha, \beta >0$,   thus $\bar{A}>0, 0<\bar{{\dom}}<1$, we observe that $\tau \leq 0$. Therefore, the equilibrium is stable if $\delta >0$. We then observe that if $\bar{{\dom}}<1/3$, then $\delta >0$. It follows that the equilibrium is stable when $\bar{{\dom}}< 1/3$. 

Now we consider the case when $\bar{{\dom}}> 1/3$. Since the equilibrium is unstable if  $\delta <0$, from (\ref{eq:delta}) we rewrite the condition $\delta <0$ equivalently as
\begin{equation}
\bar{A} <  \bar{ \mathcal{ D} }    |k|^2 \Bigg(-1+\dfrac{3\bar{{\dom}}}{\eta|k|^2+1}\Bigg), \quad \forall k.
\label{eq:Abarleq}
\end{equation}
Setting $x = \eta |k|^2$,  from (\ref{eq:Abarleq}) we infer
\begin{equation}
\bar{A}< \max_{x \geq 0}\left[ \bar{ \mathcal{ D} }   \eta^{-1}x\Bigg(-1+\dfrac{3\bar{{\dom}}}{x+1}\Bigg)\right].
\label{eq:Abarbar}
\end{equation}
To calculate the right-hand side of (\ref{eq:Abarbar}), we set the derivative of the corresponding function in $x$ equal to zero, and arrive at
\begin{equation*}
 \bar{ \mathcal{ D} }  \eta^{-1}\dfrac{-3\bar{\dom}x }{(x+1)^2} + \bar{ \mathcal{ D} }  \eta^{-1}(-1 + \dfrac{3\bar{{\dom}}}{x+1}) = 0,
\end{equation*}
\begin{equation}
x^2+2x+1-3\bar{{\dom}} = 0.
\label{eq:quadratic}
\end{equation}
We substitute the positive root   $x=-1+\sqrt{3\bar{\dom}}$ into (\ref{eq:Abarbar}) and obtain
\begin{equation}
\bar{A}< \bar{ \mathcal{ D} } \eta^{-1}(-1+\sqrt{3\bar{{\dom}}})^2.
\label{eq:stablitycondition}
\end{equation}
This together with (\ref{eq:Abarleq})   implies
(\ref{eq:unstable}) as desired.
%
\end{proof}


\section{Incorporation of Law Enforcement Agents}\label{sec:3}

In the field there is another essential component that affects the criminal behavior, namely, the presence of  law enforcement agents. We     incorporate their effects into the truncated-\textit{L\'evy}-flight model. We assume that the law enforcement agents also follow   truncated \textit{L\'evy} flights, whose  mobility parameters are possibly different than those of the criminal agents. These parameters will  determine their deployment strategy. We study the effects these law enforcement agents have on the formation of the hotspots and total number of criminal activities, and   how they depend on the mobility parameters quantitatively and qualitatively.  In Ref. \refcite{law_enforcement}, 
   only qualitative comparisons    were carried out. In Ref. \refcite{Zipkin}  law enforcement agents were also incorporated but   the focus  was on the   optimization  of the  deployment strategy through the study of   a free boundary problem.   
%
      

\subsection{Discrete model}\label{sec:disc2}

  Let $\psi_k(t)$ be the number of the law enforcement agents at site $k$ at time $t$,  and  $\tilde{A_k}(t)$ be the  attractiveness perceived by the criminals in the presence of the police agents. 
 As in Ref.  \refcite{law_enforcement}, we assume that
\begin{equation}\label{eq:tildeAk}
\tilde{A_k}(t):=e^{-\chi\psi_k(t)}A_k(t),
\end{equation}
where $\chi$ is a given constant measuring the  effectiveness of the  patrol strategy. Now we modify the model to include the effects of the   law enforcement agents. The probability   of burglarizing and moving of the     criminal agents are the same as in Section \ref{sec:overview}, except for $\tilde A$ replacing $A$. 
Thus at each time step, the system gets updated   as follows:

\textit{Step 1.} Each criminal agent decides to burglarize with probability
\begin{equation}
\tilde{p}_k(t)=1-e^{-\tilde{A}_k(t)\delta t}. 
\label{eq:pr1}
\end{equation}

\textit{Step 2.} 
 If a criminal agent chooses to commit a burglary then he will be immediately removed from the system. Otherwise he   will  move   from site $k$ to site $i$ with probability
\begin{equation}
\tilde {q}_{k \rightarrow i} (t)=\dfrac{\tilde {w}_{k \rightarrow i} (t)}{\sum\limits_{\substack{j \in \mathbb Z \\ j\neq k}}{\tilde {w} _{k \rightarrow j}(t)}}, \quad k\neq i, 
\label{eq:movq2}
\end{equation}
where
\begin{equation}
\tilde{w} _{k\rightarrow i}(t)=
\left\{\def\arraystretch{1.2}%
\begin{array}{@{}c@{\quad}l@{}}
\dfrac{\tilde{A}_i (t)}{l^{\mu}|i-k|^{\mu}}, & 1\leq |i-k| \leq L, \\
0,& \text{otherwise}.
\end{array}
\right.
\end{equation} 
 
\textit{Step 3.} 
The  law enforcement agents   move following a   truncated \textit{L\'evy} flight biased according to    the  original attractiveness field.  Hence the   probability of a law enforcement agent moving from site $k$ to site $i$ is  
\begin{equation}\displaystyle
\widehat {q}_{k \rightarrow i}(t)=\dfrac{\widehat w_{k\rightarrow i}(t)}{\sum\limits_{\substack{j \in \mathbb Z \\ j\neq k}}{\widehat w_{k\rightarrow j}(t)}}, \quad k \neq i, 
\label{eq:movq1}
\end{equation}
where
\begin{equation}
\widehat w_{k\rightarrow i}(t)=
\left\{ 
\begin{array}{lr}
\dfrac{A_i(t)}{l^{\widehat \mu}|i-k|^{\widehat{\mu}}},
 & 1\leq |i-k| \leq \widehat L, \\
0, & \text{otherwise}.
\end{array}
\right.
\label{eq:weight1}
\end{equation}
Here  $\widehat \mu \in (1, 3) $  and $\widehat L \in \mathbb N  $. These  mobility parameters of the law enforcement  agents  are not necessarily the same with    those of the criminal agents. 
We also demand that 
the total number of law enforcement agents remains a constant in time;  there is no removal or replacement of  the police agents. 
%
%

\textit{Step 4.} 
The attractiveness evolves in a way similar    to (\ref{eq:B_dis}) except for a change in the  number of burglary events in the time interval $(t, t+\delta t]$. From (\ref{eq:pr1}) we infer that there are on average $ \delta t \tilde{A_k}(t) \dom   _k(t)  $ crimes in each time interval on site $k$, and   we define the update rule as
%
\begin{equation}\label{eq:B2}
B_k(t+\delta t)=\left[(1-\eta)B_k(t)+\dfrac{\eta}{2}(B_{k-1}(t)+B_{k+1}(t))\right](1-\omega\delta t) + \theta \delta t \tilde{A_k}(t) \dom   _k(t)  ,
\end{equation}
where $\eta$, $\omega$ and $\theta$ are the same parameters as in (\ref{eq:B_dis}). 


\textit{Step 5.}   At each site a new criminal agent is  replaced with rate $\gamma$.

\begin{figure}[htpb]
\centering
\includegraphics[width=0.9\textwidth]{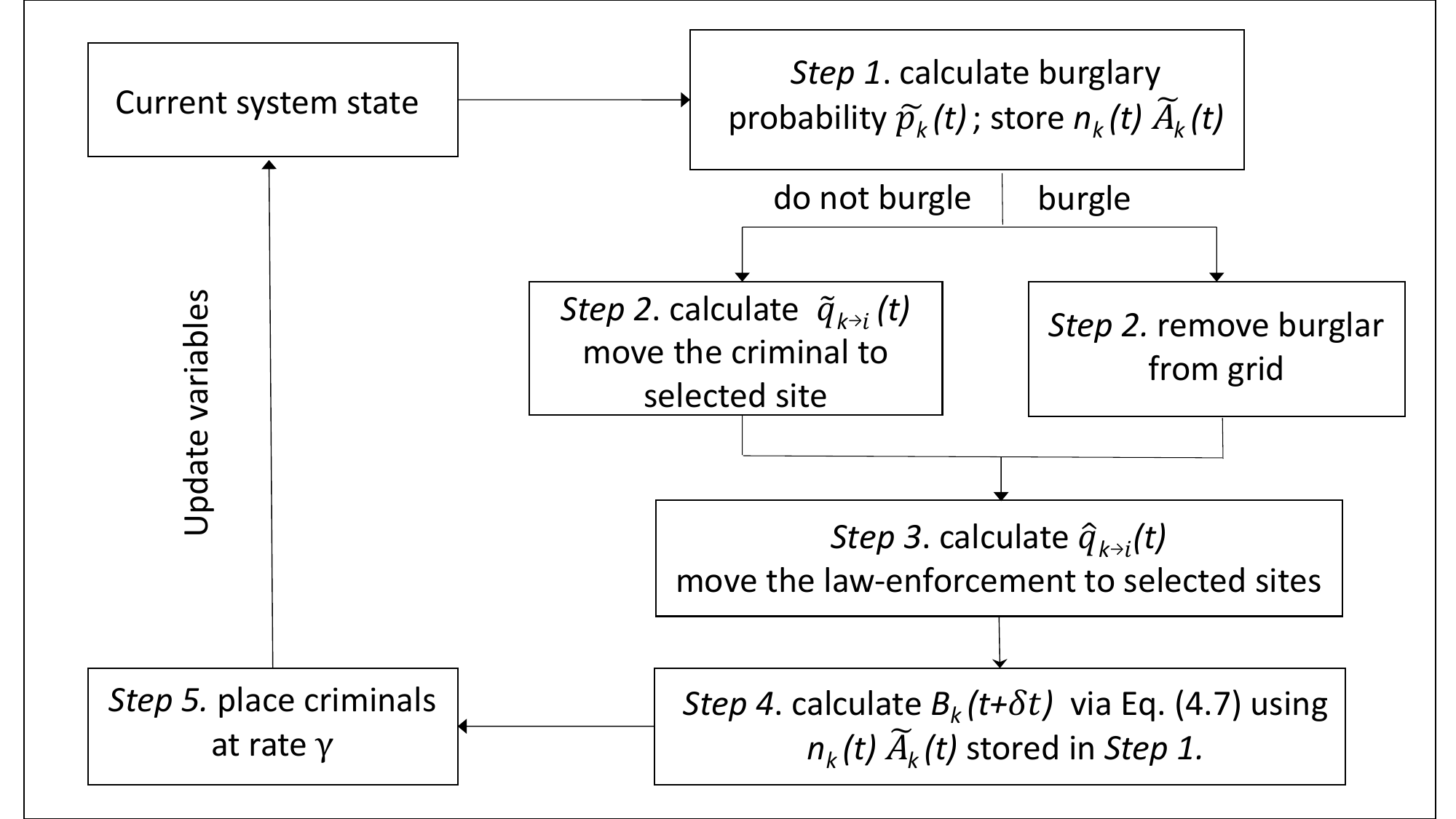}

\caption{Flowchart summarizing the discrete model with the incorporation of law enforcement agents.}\label{figdiagram2}
\end{figure}

 Figure \ref{figdiagram2} presents a visual summary of steps in the form of a flow chat. 

To conclude, the discrete mean field equation  of $(A_k (t), \tilde {A}_k (t), n_k (t),  \psi_k(t) )$ consists of (\ref{eq:tildeAk}),  (\ref{eq:B2}),  and the following equations
\begin{align}
\dom_k(t+\delta t) &=\sum_{\substack{i \in \mathbb{Z} \\ 1 \leq |i-k| \leq L}}{[1-\tilde{A}_i(t)\delta t]\dom_{i}(t) \tilde{q}_{i \rightarrow k}(t)}+\gamma\delta t, \label{eq:rhoDisc22} \\
\psi_k(t+\delta t)&=\sum_{|i-k|\leq \widehat  L}{\psi_i(t)\widehat q_{i \rightarrow k}(t)}. 
\label{eq:BTLFP}
\end{align}


\subsection{Continuum model}\label{sec:cont}

The derivation of the continuum equations for the attractiveness field and the criminal number distribution  is very similar as that in  Sec.  \ref{continulimit}; basically, from (\ref{eq:B2}) and (\ref{eq:rhoDisc22}) we obtain   (\ref{eq:contA}) and (\ref{eq:contrho}) with $A$ replaced by $\tilde A$ when suitable:
%
\begin{align}
\dfrac{\partial A}{\partial t}&=\dfrac{ l^2\eta}{2 \delta t}A_{xx}-\omega(A-A^0)+\theta \dom \tilde {A}, 
\label{eq:contA2}\\
\dfrac{\partial \dom}{\partial t}&=\mathcal D   \vec{\nabla} \cdot \left [ \vec{\nabla} \dom - \dfrac{2 \dom}{
\tilde A} \vec{\nabla}\tilde A \right]-\tilde{A}\dom+\gamma.
\label{eq:contrho23}
\end{align} 
This however will not lead to the identical system since now (\ref{eq:contA2}) and (\ref{eq:contrho23}) are part of a larger system  which also includes the dynamics of  the component of law enforcement agent. 
 With a similar derivation as   in   Sec.  \ref{continulimit}, from (\ref{eq:BTLFP}) we obtain
\begin{equation}\label{eq:conl}\displaystyle
\dfrac{\partial \psi}{\partial t}=
  \widehat  { \mathcal{ D} }   \vec{\nabla} \cdot \left [ \vec{\nabla} \psi - \dfrac{2 \psi}{A} \vec{\nabla}A \right],  
\end{equation}
where   
\begin{equation}\label{eq:D12}
\widehat  { \mathcal{ D} }=\dfrac{l^2  }{\delta t }\dfrac{z^{\ast}_{\widehat \mu, \widehat L}}{z_{\widehat \mu, \widehat{L}}}. 
\end{equation}

To conclude, 
the continuum model with law enforcement effects  consists of (\ref{eq:tildeAk}) and  (\ref{eq:contA2})-(\ref{eq:conl}). 
 %
%

\subsection{Computer simulations}\label{sec:cont22}
In order to verify the validity of our continuum model,
  and to compare results with the discrete model, 
  we perform direct numerical simulations. 
We consider the basic deployment strategies including    a biased random walk (BRW)   and a truncated \textit{L\'evy} flight (TLF) with the same mobilities as those of the criminal agents. 
 %
 As in Ref. \refcite{law_enforcement}, we also include the     base case where  the law enforcement agents patrol random routes, that is, an unbiased random walk (URW). Here the law enforcement agents do not focus their attention on any particular place. 
 In this case, the continuum equation for the dynamics of law enforcement agents is just the unbiased Brownian motion. \cite{levycrime}

To implement the discrete model,  we consider a lattice grid     on a spatial domain  $[0, 1]$  with the lattice spacing being ${l}=1/60$. The computation assumes periodic boundary conditions.
The algorithm  used for the continuum simulation is very similar to that used  in Section \ref{sec:comparisons}. 
Particularly,  we   use a semi-implicit time discretization, with the time-stepping algorithms as follows:
\begin{align}
\tilde{A}^{(m)}&=A^{(m)} e^{-\chi \psi^{(m)}},\\
 A^{(m+1)}& =A^{(m)}+\Delta t\left (\eta A_{xx}^{(m)}-A^{(m)}+A^0+n^{(m)} \tilde{A}^{(m)}\right),\\
 n^{(m+1)}& =n^{(m)}+\mathcal{D}\Delta t\left[n^{(m)}_{xx}-\left (\dfrac{2n\tilde{A}_x^{(m+1)}}{\tilde{A}^{(m+1)}}\right)_x \right]+\Delta t \left(-\tilde{A}^{(m+1)}n^{(m)}+\gamma \right),\\
 \psi^{(m+1)}& =\psi^{(m)}+\hat{\mathcal{D}}\Delta t \left[\psi^{(m)}_{xx}-\left (\frac{2\psi^{(m)} A^{(m+1)}_x}{A^{(m+1)}} \right)_x \right].
\end{align}
Here  $f^{(m)}$ represents a quantity $f$ at $m$th time step. 
To discretize the  functional space  of the solutions, we use the fast Fourier transform (FFT).

  Figs.  \ref{fig:rw},  \ref{fig:brw},  and \ref{fig:btlf} below show the discrete and the continuum models corresponding to the three deployment strategies. Good agreement is observed  in all cases, which  validates the continuum models. 
   It is expected that  the  unbiased random walk     in Fig.  \ref{fig:rw}  does  not reduce hotspot activity and is the least effective of all the three strategies,  which   agrees  with the empirical evidence in Ref. \refcite{law_enforcement}.   However the comparison  of Fig. \ref{fig:brw}  and Fig. \ref{fig:btlf}     is less trivial. 
It seems that 
%
Fig. \ref{fig:btlf}   shows  higher deployment effectiveness, as   a steady state is reached faster. 
However for a better comparison we need  to  first quantify the  effectiveness of the deployment strategies.

\begin{figure}[ht]
\centering 
\begin{subfigure}[b]{0.455\textwidth}
\centering
\includegraphics[width=1  \textwidth]{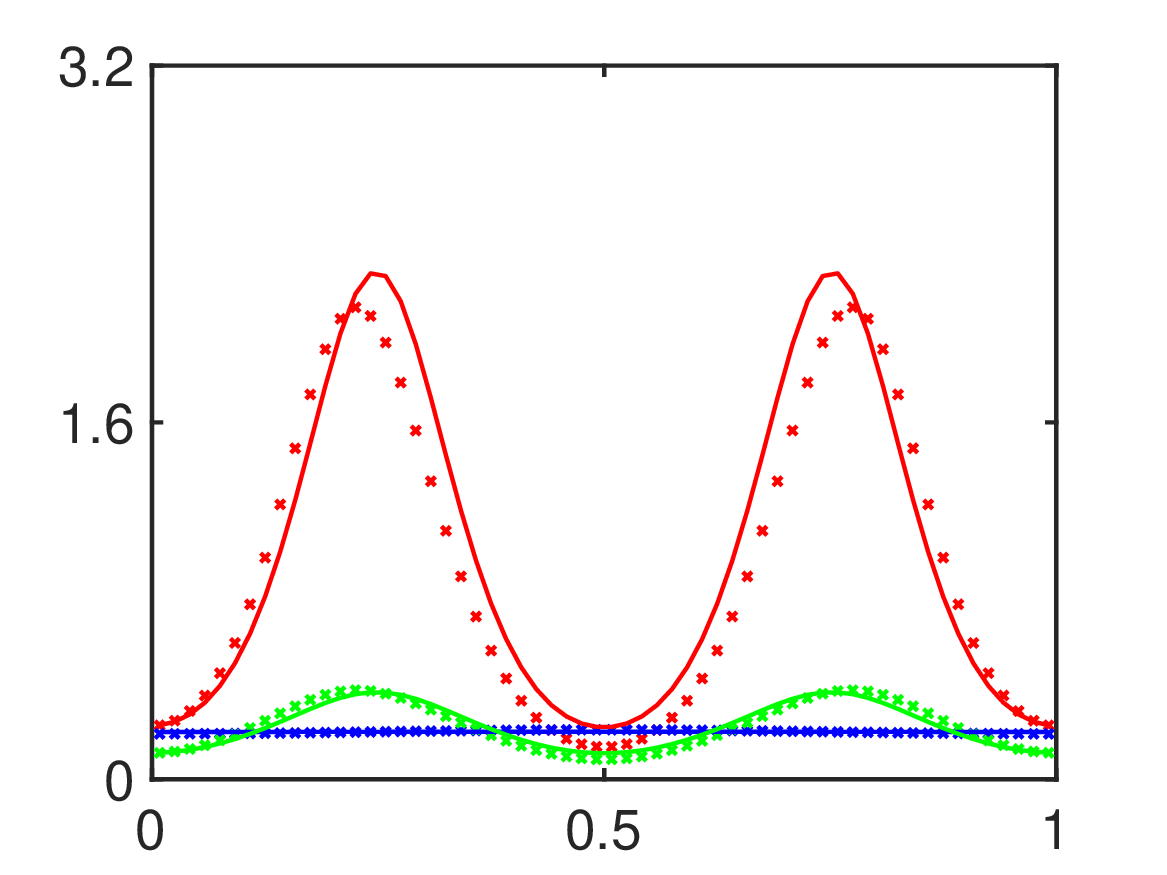}
\caption{$t = 5$}
\label{fig:brw_1_1}
\end{subfigure}\hskip -4.3 mm
\begin{subfigure}[b]{0.525\textwidth}
\centering
\includegraphics[width=1 \textwidth]{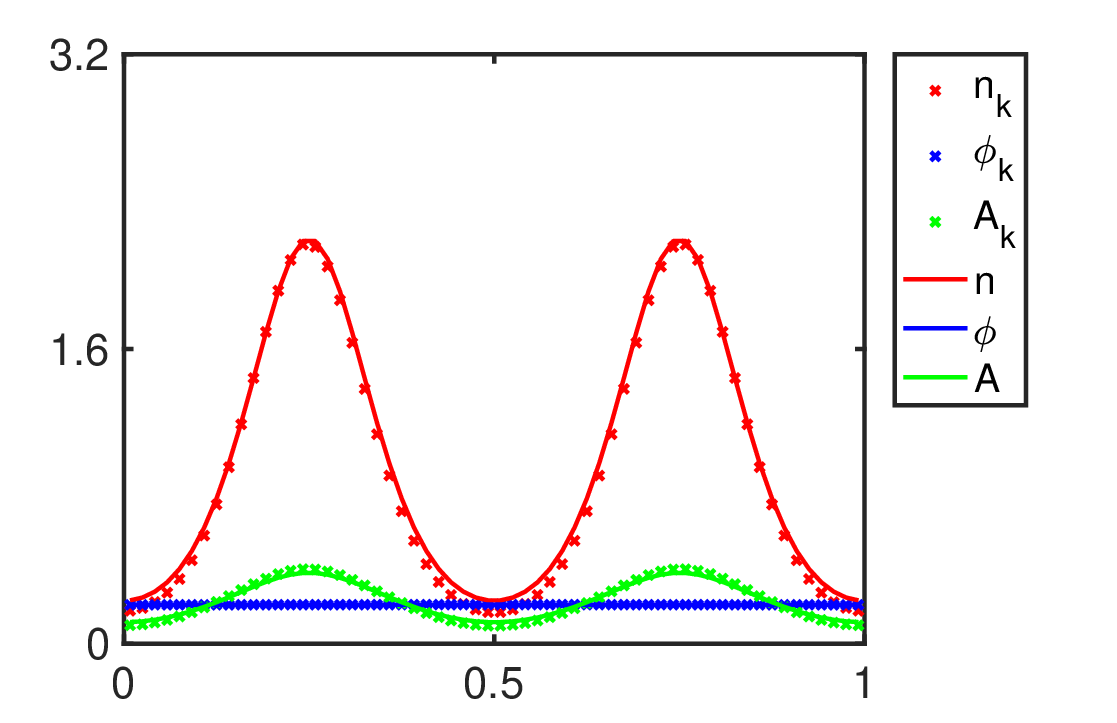}
\caption{$t = 10$}
\label{fig:brw_5_1}
\end{subfigure}
 \vspace*{-3pt}
\caption{Comparisons of the discrete   and  the continuum models with the unbiased-random-walk deployment strategy.  The discrete model  (\ref{eq:tildeAk})-(\ref{eq:B2})  is shown with  cross lines, and the continuum model  (\ref{eq:tildeAk}) and (\ref{eq:contA2})-(\ref{eq:conl})  is shown with  solid lines. 
The   attractiveness field,   the criminal  and  the law enforcement agent number distributions    are shown  with    green, red and blue  lines.    The initial conditions (at $t=0$) are taken to be  $\psi  = 1/3\sin(\pi x)$,       $B \equiv 0$ and ${\dom} =1-0.3\cos (4 \pi x)$. Parameters are $\chi=8$, $A^0  = 1-0.5\cos(4\pi x)$,   $l = 1/60$, $\delta t = 0.01$, $L =9$,  ${\eta} = 0.1$, $\gamma = 0.3$, $\omega = 1$, $\theta =1$,   and $\widehat L=1$. The system enters a steady state   at roughly $t=10$. }
\label{fig:rw}
\end{figure}
\begin{figure}[ht]
\centering
\begin{subfigure}[b]{0.455\textwidth}
\centering
\includegraphics[width=1  \textwidth]{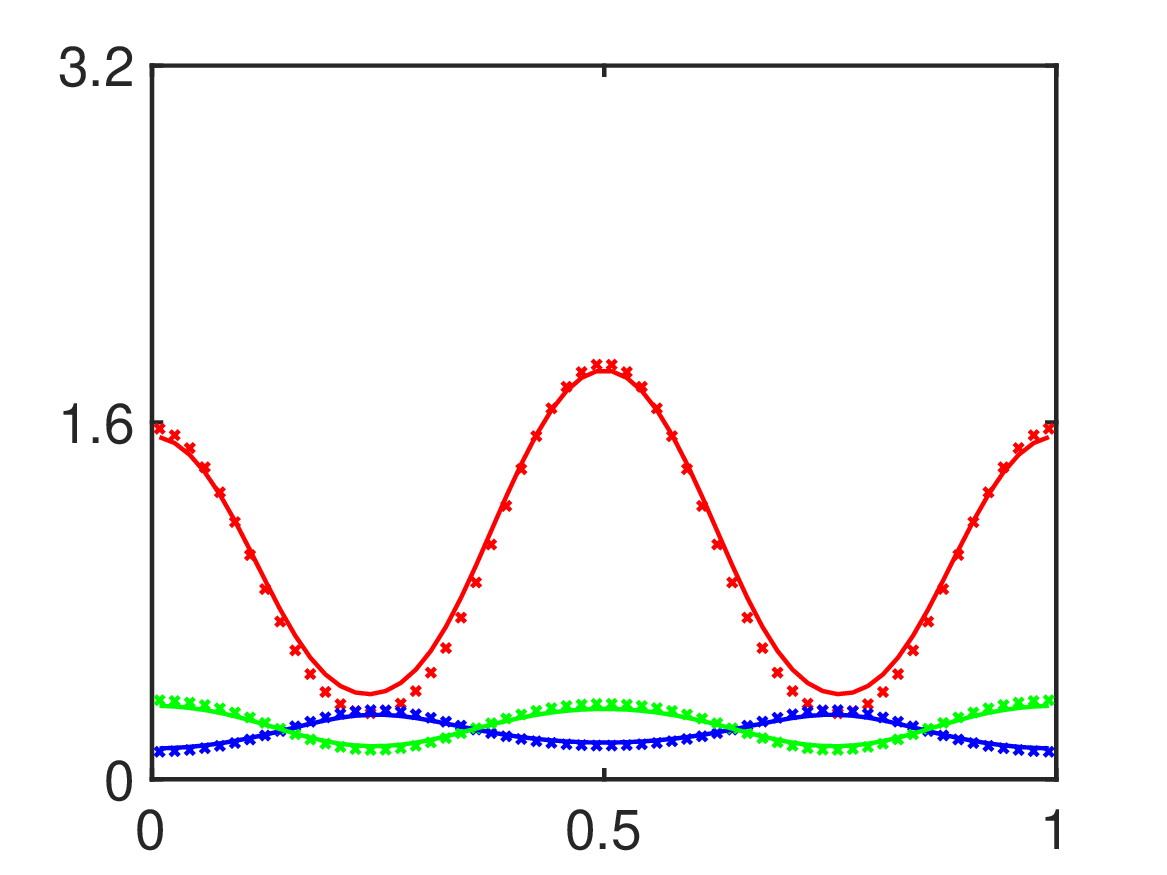}
\caption{$t = 5$}
\label{fig:brw_1_2}
\end{subfigure}
\hskip -4.3 mm
\begin{subfigure}[b]{0.525\textwidth}
\centering
\includegraphics[width=1 \textwidth ]{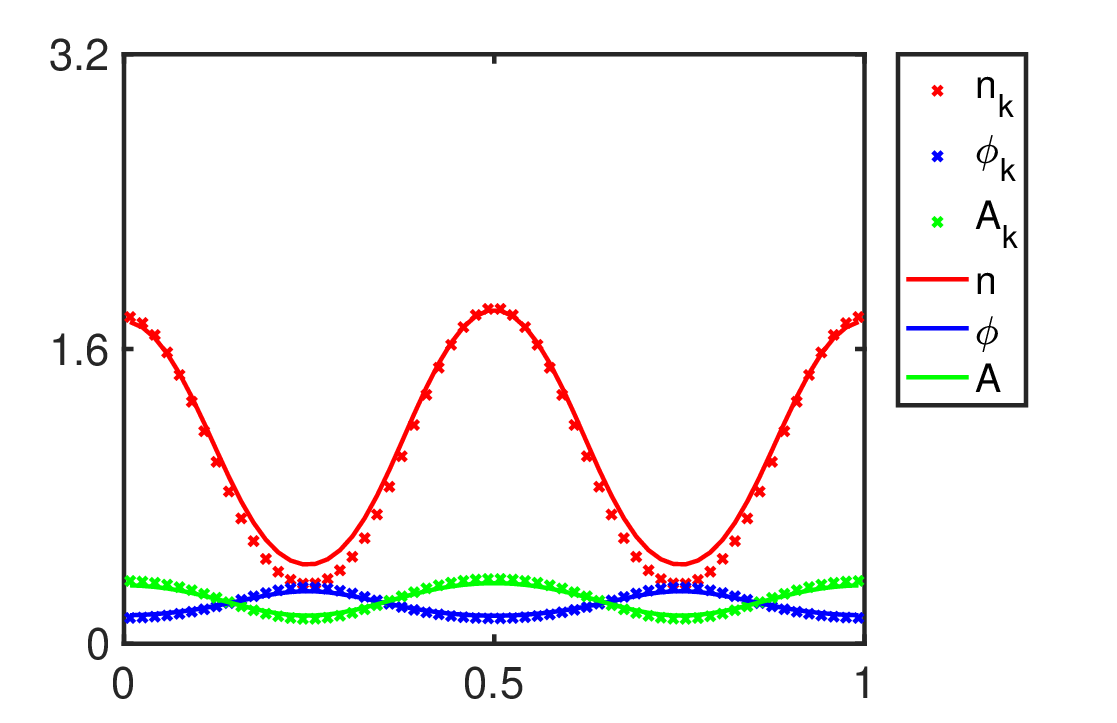}
\caption{$t = 10$}
\label{fig:brw_5_2}
\end{subfigure}
 \vspace*{-3pt}
\caption{Comparisons of the discrete  and  the continuum models with a biased-random-walk deployment strategy. 
 All the   parameters and data are the same as in Fig. \ref{fig:rw}. The system enters a steady state roughly   at  $t=10$.  
}
\label{fig:brw}
 \end{figure}
  \begin{figure}[ht]
 \centering
\begin{subfigure}[b]{0.455\textwidth}
\centering
\includegraphics[width=1  \textwidth]{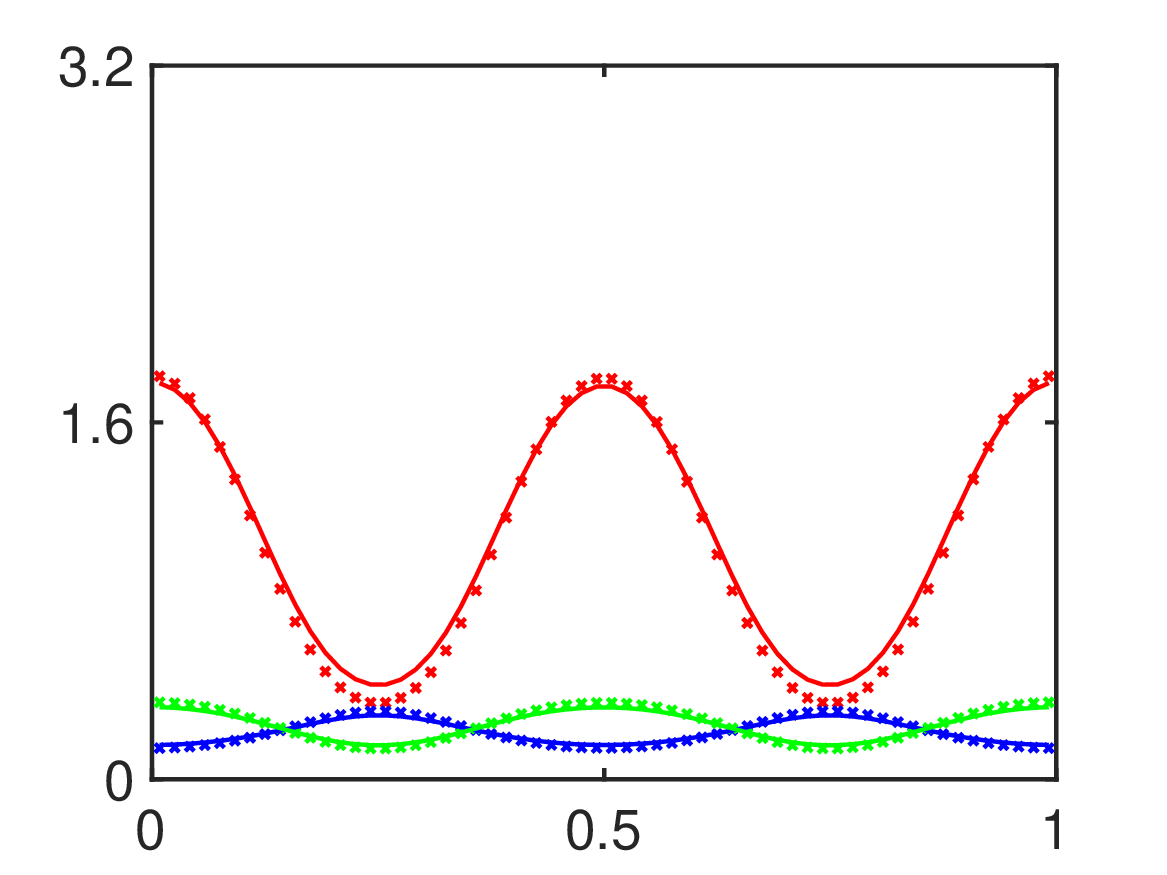}
\caption{$t = 5$}
\label{fig:btlf_1}
\end{subfigure}
\hskip -4.3 mm
\begin{subfigure}[b]{0.525\textwidth}
\centering
\includegraphics[width=1  \textwidth]{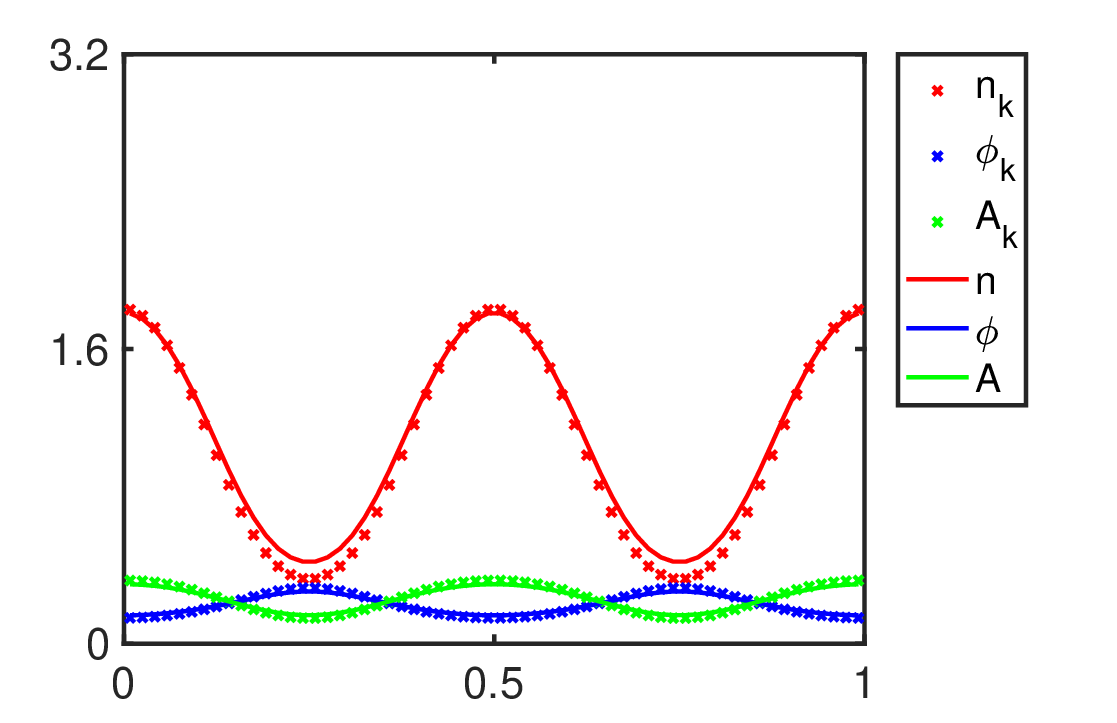}
\caption{$t = 10$}
\label{fig:btlf_5}
\end{subfigure}
\caption{Comparisons of  the discrete  and  the continuum models when  the deployment strategy of a truncated \textit{L\'evy} flight is adopted. 
Here    $\widehat L=L=9$, $\widehat \mu = \mu=2.5$,  and  all the  other parameters and data are the same as in Fig. \ref{fig:rw}. The system enters a steady state roughly at   $t=5$.  
}
\label{fig:btlf}
\end{figure}

   \begin{remark} \label{rmk9}
    In  Ref. \refcite{law_enforcement},  a "peripheral interdiction" was  also considered,  which sends the police to the perimeters of the crime hotspots instead of the centers. However this  deployment strategy is not considered  here, as  the criminals  can take long jumps now, and protecting the perimeters of a hotspot no longer necessarily prevents them from entering the center. 
   \end{remark}

\subsection{Quantitative comparisons of the patrol effectiveness}\label{sec:evaluation}
  
We  compute    the  cumulative number of burglaries for the system. 
%
%
  %
%
%
 %
From (\ref{eq:pr1}) we infer that   the total expected number of burglary events over the whole domain  up to time $T$   
  equals to $\sum_{k}\sum_{t=n \delta t, \,\,\, 0<t <T}{\tilde{A_k}(t) \dom_k (t) \delta t}$, where $k$ is  sum  over all the grid points in the
domain. When the domain size is kept fixed and $l$ is sent to zero, the
total number of grid points in this domain will increase to infinity.
Therefore, to make sense of the above quantity, we rescale it by
multiplying it with $l$, and physically it means the    averaged expected 
number of burglaries.  Then
taking the limit  as $\delta t$ and $l$ become small, the rescaled double sum yields a double integral, denoted as $S(T)$:
\begin{equation}\label{eq:st}
S(T) =  \int_{0}^{T}\int_{\mathcal{M}}{\tilde{A}(x,t){\dom}(x,t)dxdt}, 
\end{equation}where $\mathcal{M}$ denotes the  spatial domain on which the lattice grid lives. 
The instantaneous global crime  rate  can be defined
 as 
\begin{equation}
R(t):=\dfrac{\partial S(t)}{\partial t}=\int_{\mathcal{M}}{\tilde{A}(x,t){\dom}(x,t)dx}.
\label{eq:rt}
\end{equation}
Fig.  \ref{fig9}  below shows the simulations of $R(t)$  and $S(t)$ when zero law enforcement agent is put in the system, and when  one of  the  deployment strategies mentioned in Section \ref{sec:cont22} above are employed, that is,  an unbiased and a   biased random walk,  and  a truncated \textit{L\'evy} flight with the same mobility parameters as the criminals. 
%
 %
%

\begin{figure}[htpb]
\centering
\begin{subfigure}[b]{1\textwidth}
 \hskip 8 mm  \includegraphics[width = 0.8\textwidth]{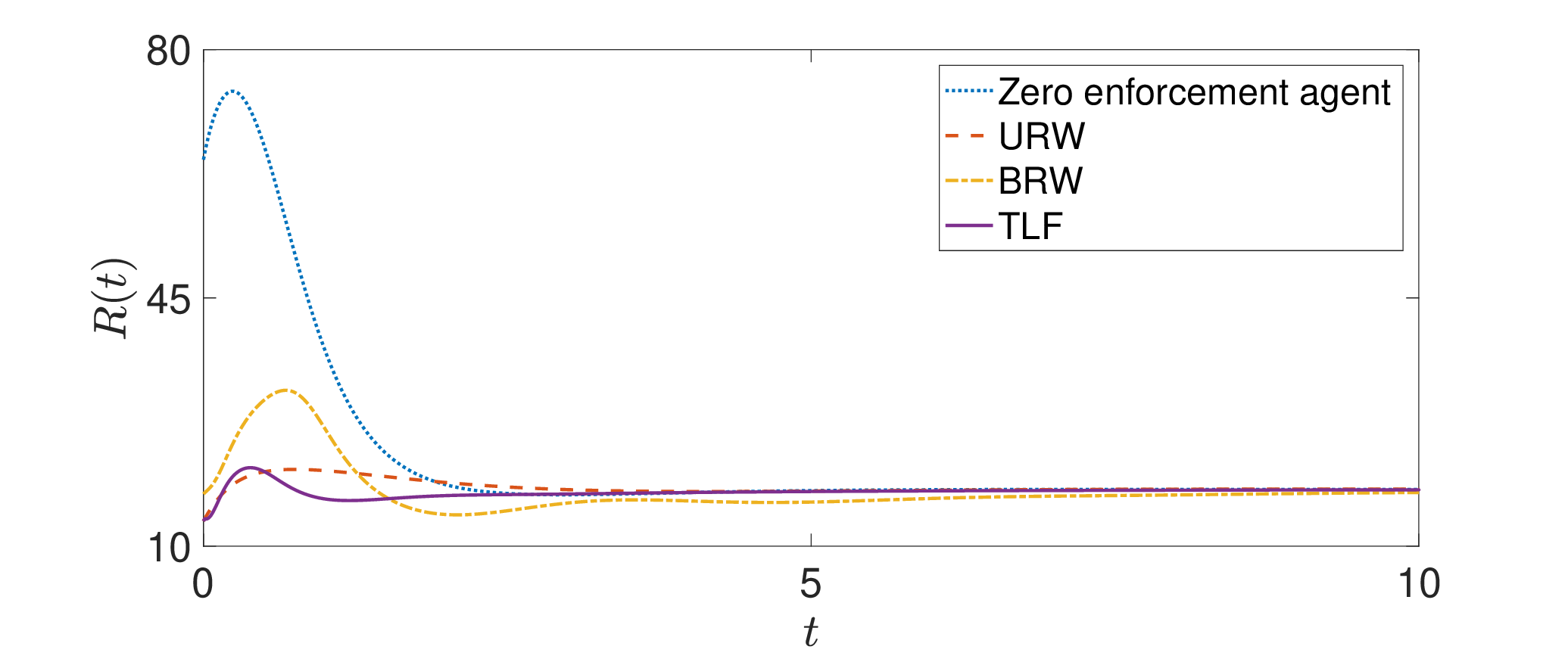}
  \end{subfigure}
  \centering
\begin{subfigure}[b]{1\textwidth}
 \hskip 8 mm  \includegraphics[width =0.80\textwidth]{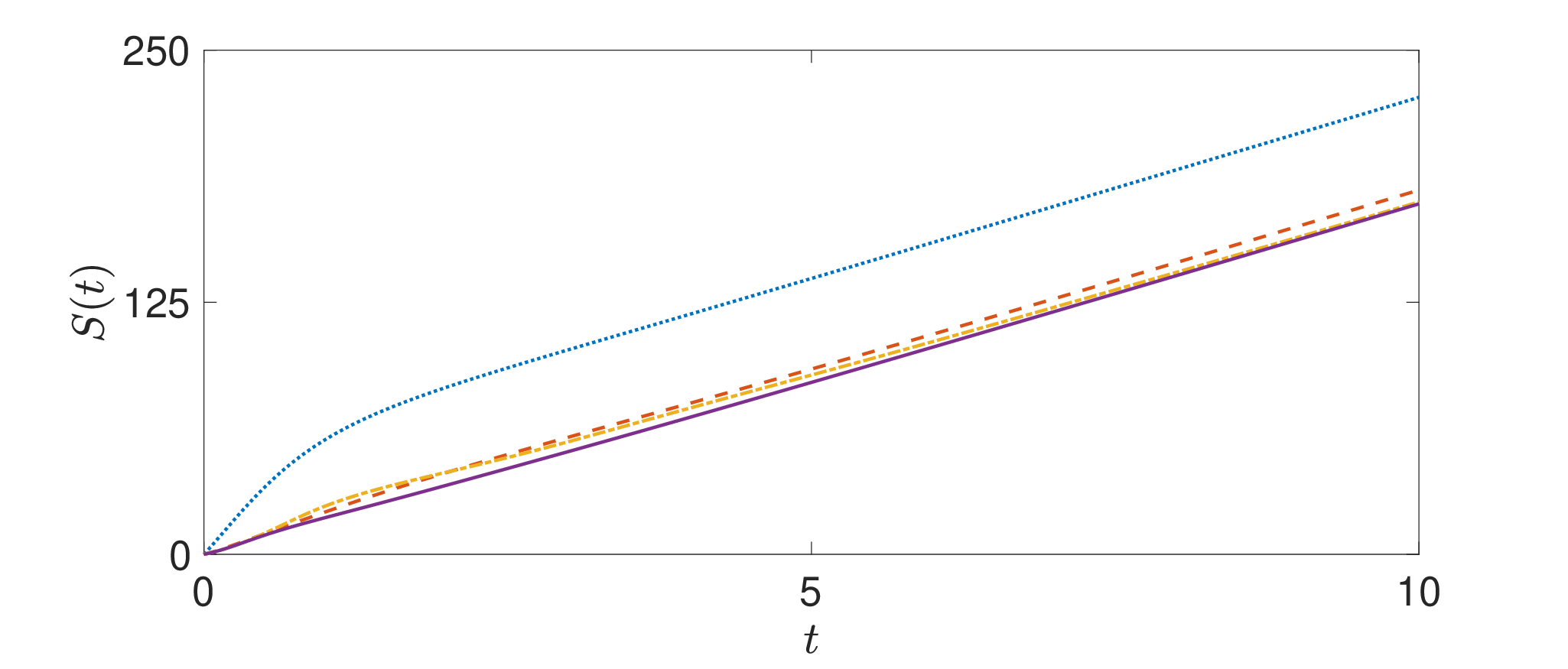}
\end{subfigure}
\caption{Output of     $R(t)$     and   $S(t)$, when  there is no law enforcement agent, and when the deployment strategy of an unbiased random walk (URW), a biased random walk (BRW), or a truncated \textit{L\'evy} flight (TLF) is adopted. 
Here all the    parameters and data  are the same as in Figs. \ref{fig:rw}-\ref{fig:btlf}, except for $l=1$ and the spatial domain $\mathcal M=[0, 60]$.}
\label{fig9}
\end{figure}

We observe   that 
    $R(t)$  approaches a constant steady state   independent of the incorporation of law enforcement agents. In fact, we find that this steady state only     depends    on the rate of criminals entering the system and the size of the domain:
\begin{theorem}\label{thm:rt}
Once the system (\ref{eq:tildeAk}) and  (\ref{eq:contA2})-(\ref{eq:conl})   is in a steady state at time $T$,  we have 
\begin{equation}\label{eq:m}
R(t)
 = \gamma |\mathcal{M}|, \quad \forall t \geq T, 
\end{equation} where $|\mathcal{M}|$ is the size of the domain $\mathcal M$. 
\end{theorem}
\begin{proof}
We integrate (\ref{eq:contrho23}) over the domain $\mathcal{M}$
\begin{align}
\dfrac{\text{d}}{\text{d}t}\int_{\mathcal{M}} {\dom}\, dx & =\int_{\mathcal{M}}{{\dom}_t} \, dx \notag \\
&= { \mathcal{ D} }   \int_{\mathcal{M}}  \left[ \tilde{A} \left(\dfrac{{\dom}}{\tilde{A}}\right)_{xx} - \dfrac{{\dom}}{\tilde{A}}\tilde{A}_{xx} \right] - \tilde{A}{\dom} + \gamma  \, dx \notag \\
&= { \mathcal{ D} } \left [\tilde{A}(\dfrac{{\dom}}{\tilde{A}})_x-\dfrac{{\dom}}{\tilde{A}}\tilde{A}_x\right] \bigg|_{\mathcal{M}}-\int_{\mathcal{M}}{(\tilde{A}{\dom}-\gamma) \, dx} \notag \\
 & = \gamma |\mathcal{M}| - \int_{\mathcal{M}}\tilde{A}{\dom} \, dx. \label{nonhomogenous}
\end{align}
Here the periodic boundary conditions are applied.  
When the system is at a steady state,  the left-hand-side of (\ref{nonhomogenous}) vanishes, 
 and hence (\ref{eq:m}) follows
as desired. 
\end{proof}

In the original  random-walk model \cite{uclaModel}   the crime suppression was built-in to the decay of the attractiveness. 
This was used to model the finite lifetime of the repeat victimization effect. 
Here we add additional law enforcement who  curb the crimes by decreasing the attractiveness. We noted that in Fig. \ref{fig9}   with or without law enforcement agents the steady state  crime rate   is identical. 
This happens essentially because of the  constant  replacement rate, and was first observed in the original random-walk model. \cite{uclaModel}
  Nevertheless,  Fig. \ref{fig9} shows that  law enforcement agents do affect the  number of burglary events  cumulated before the crime rate enters the steady state.  
Thus it seems reasonable to  measure the deployment efficiency using $S(T)$ at the time $T$, when $R(T)$ just enters  the equilibrium.   
In  Fig.  \ref{fig9},   $T$ can be chosen as $5$, as 
 $R(T)$ is always  within negligible difference  from the steady state crime rate  after time $5$. 
 %
 
 Tables 1, 2 and 3 below display $S(5)$ for the three deployment strategies shown in Fig. \ref{fig9} (the system enters the steady state roughly at $T=5$ in all three cases). 
 We    observe from these tables that     the truncated  \textit{L\'evy} flight  with the same mobility parameters as the criminals  is the most effective deployment strategy in terms of reducing the total number of crime events. This quantitative result coincides with our intuition and the qualitative comparisons in Figs. \ref{fig:rw}-\ref{fig:btlf}.

\begin{table}[ht]
\caption{Comparisons of the global   cumulative number of the burglary events till  time $T = 5$ between different deployment strategies. `Improvement I'  shows the   improvement compared to the base case with zero enforcement agents.  `Improvement II' corresponds to the improvement compared to the  unbiased-random-walk deployment strategy. The  parameters and initial conditions are the same as in Fig.  \ref{fig9}. In this case
there are initially two regions of high attractiveness.}
\footnotesize
\centering
{\begin{tabular}{@{}cccc@{}} 
\toprule
 The Police Deployment Strategy & S(5) & Improvement I& Improvement II\\
 \colrule
 Zero Law Enforcement Agent & 137 & - & -\\
 Unbiased Random Walk & 91.87 & 32.94$\%$ & - \\
 Biased Random Walk & 88.91 & 35.1$\%$ & 3.22$\%$  \\
   Truncated \textit{L\'evy} Flight & 85.26 & 37.76$\%$ & 7.19$\%$ \\
 \botrule
\end{tabular}}
\label{tab:E_table}
\end{table}

\begin{table}[ht]
\caption{Comparisons of the global  cumulative number of the burglary events till time $T = 5$  between different deployment strategies. Here the initial conditions (at $t=0$) are taken to be ${\dom}=1-0.3\cos(8\pi x)$ and $A^0=1-0.5\cos(8\pi x)$, and all the other parameters and data are the same as in Table 1. In this case, there are four regions of high attractiveness initially.} 
\centering
\footnotesize
{\begin{tabular}{ @{}cccc@{} }
 \toprule
 The Police Deployment Strategy & S(5) & Improvement I& Improvement II\\
 \colrule
 Zero Law Enforcement Agent & 136.46& - & -\\
Unbiased Random Walk & 97.83 & 28.3$\%$ & - \\
 Biased Random Walk & 85.69 & 37.2$\%$ & 12.41$\%$  \\
   Truncated \textit{L\'evy} Flight & 82.73 & 39.37$\%$ & 15.44$\%$ \\
 \botrule
\end{tabular}}
\label{tab:E_table_8}
\end{table}

\begin{table}[ht]
\caption{Comparisons of the global  cumulative number of the burglary events till  time $T = 5$ between different deployment strategy.  Here the initial conditions are taken to be ${\dom}=1-0.3\cos(16\pi x)$ and $A^0=1-0.5\cos(16\pi x)$, and all the other parameters and data are the same as in Table 1. In this case,  there are  eight regions of high attractiveness initially.}
\centering
\footnotesize
{\begin{tabular}{ @{}cccc@{} }
 \toprule
The Police Deployment Strategy & S(5) & Improvement I& Improvement II\\
 \colrule
 Zero Law Enforcement Agent & 136.4 & - & -\\
Unbiased Random Walk & 102.74 & 24.68$\%$ & - \\
 Biased Random Walk & 83.11& 39.07$\%$ & 19.1$\%$  \\
   Truncated \textit{L\'evy} Flight & 79.2 & 41.93$\%$ & 22.91$\%$ \\
\botrule
\end{tabular}}
\label{tab:E_table_16}
\end{table}

 %


 \section{Conclusion}
 
 In this paper, we apply the truncated \textit{L\'evy} flights to the class of agent-based crime models for residential burglary. The truncation becomes a  parameter  that restricts the mobility of the agents. We study both the discrete model and its continuum limit which agree very well in  computer simulations. We find that   the  continuum system behaves like modified Brownian  dynamics. This indicates that the continuum version of the original random-walk model in Ref. \refcite{uclaModel}, which also has a Brownian dynamics, can be utilized here with a modified diffusion coefficient. For instance,  the stability analysis in the original paper \cite{uclaModel} can be modified and applied   to our model efficiently. This serves as a first step towards the weakly nonlinear analysis and bifurcation theory which can help  law-enforcement  to
understand the feedback between treatment and    hotspot dynamics. \cite{shortPNAS, short2010}.
 Then we examine  the impact of introducing police into the truncated-\textit{L\'evy}-flight model, whose mobility parameters determine the deployment strategies. We observe that the    strategies can affect the global cumulative number of the burglary events   before the  system steady state is reached. We make a quantitative comparison of the deployment strategy efficiency accordingly. We   find that 
      the truncated \textit{L\'evy} flight   with the same mobility parameters as those of the criminal agents is  the most efficient, compared to the deployment strategies of  an unbiased and a biased random walk.

 For the future work, on the one hand, we can extend the truncated-\textit{L\'evy}-flight model to two dimensional-space, which is more realistic when modeling   household distributions in typical urban area. Then we can explore whether the ``finite size effects'' observed previously in the original model  \cite{uclaModel} is also an attribution of our model, namely, whether the transience of the hotspot dynamics in the discrete simulations will depend on the initial   population size.  On the other hand, we can continue the study of  the dependence of the law enforcement patrol efficiency upon the mobility parameters of the agents. A complete parametrization of the efficiency with the mobility parameters  may be suggestive for  the police patrol strategy design.

\section*{Acknowledgment}
 
We would   like to thank    Theodore Kolokolnikov,   Martin Short,  Scott McCalla, Sorathan Chaturapruek,     and  Adina Ciomaga for helpful discussions.  This work is supported by NSF grant DMS-1045536, NSF grant DMS-1737770,  and ARO MURI grant W911NF-11-1-0332. L. W. is also partly supported by NSF grant DMS-1620135. 

\renewcommand\bibname{References}

\end{document}